\documentclass[review]{elsarticle}

\usepackage{lineno,hyperref}
\usepackage{graphicx}
\usepackage{amsmath}
\usepackage{svg}
\usepackage{pdfpages}
\usepackage[colorinlistoftodos]{todonotes}

\usepackage{wrapfig}
\usepackage{tabularx}

\usepackage{subfig}
\usepackage{booktabs}

\usepackage[vpos=.92\paperheight,angle=0,fontsize=12pt]{draftwatermark}
\SetWatermarkText{\href{https://doi.org/10.1016/j.neucom.2021.05.081}{Published article -- DOI: 10.1016/j.neucom.2021.05.081}}

\newcommand*\modified{}

\modulolinenumbers[5]

\journal{Elsevier Neurocomputing}

\begin{document}

\begin{frontmatter}

\title{Explaining Clinical Decision Support Systems in Medical Imaging using Cycle-Consistent Activation Maximization}%

\author[mymainaddress,mysecondaryaddress]{Alexander Katzmann\corref{mycorrespondingauthor}}
\ead{alexander.katzmann@siemens-healthineers.com}
\author[mymainaddress]{Oliver Taubmann}
\author[mymainaddress]{Stephen Ahmad}
\author[mymainaddress]{Alexander M\"uhlberg}
\author[mymainaddress]{Michael S\"uhling}
\author[mysecondaryaddress]{Horst-Michael Gro\ss}

\cortext[mycorrespondingauthor]{Corresponding author}

\address[mymainaddress]{Siemens Healthineers, Computed Tomography, 91301 Forchheim, Germany}
\address[mysecondaryaddress]{Ilmenau, University of Technology, Neuroinformatics and Cognitive Robotics Lab, 98693 Ilmenau, Germany\newline\newline Published article  \href{https://doi.org/10.1016/j.neucom.2021.05.081}{DOI: 10.1016/j.neucom.2021.05.081}\vspace{-2em}}

\begin{abstract}
		Clinical decision support using deep neural networks has become a topic of steadily growing interest. While recent work has repeatedly demonstrated that deep learning offers major advantages for medical image classification over traditional methods, clinicians are often hesitant to adopt the technology because its underlying decision-making process is considered to be intransparent and difficult to comprehend. In recent years, this has been addressed by a variety of approaches that have successfully contributed to providing deeper insight. Most notably, additive feature attribution methods are able to propagate decisions back into the input space by creating a saliency map which allows the practitioner to ``see what the network sees.'' However, the quality of the generated maps can become poor and the images noisy if only limited data is available---a typical scenario in clinical contexts. We propose a novel decision explanation scheme based on CycleGAN activation maximization which generates high-quality visualizations of classifier decisions even in smaller data sets. We conducted a user study in which {\modified{}we evaluated our method on the LIDC dataset for lung lesion malignancy classification, the BreastMNIST dataset for ultrasound image breast cancer detection, as well as two subsets of the CIFAR-10 dataset for RBG image object recognition. Within this user study, our method clearly outperformed existing approaches on the medical imaging datasets and ranked second in the natural image setting.} With our approach we make a significant contribution towards a better understanding of clinical decision support systems based on deep neural networks and thus aim to foster overall clinical acceptance.
\end{abstract}


\end{frontmatter}

\section{Introduction}
\label{Introduction}
Within the last years, clinical decision support using deep neural networks (DNNs) has become a topic of steadily growing interest. This includes applications in microscopy and histopathology \cite{coudray2018classification,bayramoglu2016deep}, time-continuous biosignal analysis \cite{ganapathy2018deep,yuan2017wave2vec}, and, quite prominently, medical image analysis for volumetric imaging data as generated by computed tomography \cite{hua2015computer,wurfl2016deep}, positron emission tomography \cite{dong2020deep,ding2019deep} or magnetic resonance imaging \cite{shen2017deep,trebeschi2017deep,liu2018deep}.
In the field of medical imaging, recent work has demonstrated a variety of applications for DNNs, such as organ segmentation \cite{gibson2018automatic}, anomaly detection \cite{chalapathy2019deep}, lesion detection \cite{yan2018deeplesion}, segmentation \cite{kamnitsas2017efficient} and assessment \cite{hosny2018deep}, providing major advantages and even repeatedly outperforming gold-standard human assessment \cite{liu2019comparison}.

A nearby field of similarly growing research interest established with the publications of Kumar et~al.~and Aerts et~al.~\cite{kumar2012radiomics,aerts2014decoding} is {\modified ``r}adiomics'' using traditional machine learning (ML) techniques. Compared to deep learning techniques, traditional ML methods like random forests and support vector machines have a largely transparent decision-making process, which is generally easier to comprehend and/or depict -- a clear argument for their preference in clinical practice.
Many publications have shown the advantages of DNNs in comparison to traditional machine learning techniques, such as the ability to learn descriptive features from data instead of a complex and expensive handcrafted feature design, as well as an improved classification performance on medical imaging tasks \cite{litjens2017survey,katzmann2018predicting}, with some architectures being on par with gold-standard human assessment \cite{liu2019comparison}. However, as DNNs learn features from the given data, the semantic of these features is in general not immediately evident. Thus, clinicians understandably approach these methods with a high degree of skepticism.

\subsection{Related Work}
A variety of methods have been proposed for decision explanation in DNNs{\modified{}, and various categorizations for these have been presented. Linardatos et al.~\cite{linardatos2021explainable} suggested to categorize different approaches with respect to four main determinants, namely the provision of local vs.~global explanations, the modality of classified data (images, text, graphs, \ldots), the purpose of explanation (e.\,g.~black-box explanation, model simplification, etc.) and the attribute of being model-specific vs.~model agnostic. For the sake of simplicity, we follow the notion of a method being model-agnostic if it can be applied without requiring a specific type of network architecture. Within this work we will focus on black-box, local explanations of image classifications.} %

Common to these approaches is a visualization of the decision process, or part of it, by highlighting regions in the input image which the algorithm identified as decision-relevant. While the  provided insight is limited to the input space, rather than the inner workings of the network, it allows a human observer to visually inspect the classifiers's regions of interest and whether these match with the expected image areas.

Model-{\modified{specific}} analysis methods like DeconvNet~\cite{noh2015learning}, (Grad)CAM(++) \cite{zhou2015cnnlocalization,selvaraju2017grad,chattopadhay2018grad} and Attention Gated Networks~\cite{schlemper2019attention} work by explicitly modifying the network structure in such a way that there is an immediate spatial representation of relevance as a condition for a successful classification, e.\,g.~by predicting attention areas and masking activations in regions which are expected by the network to be non-relevant. Typically, methods like these {\modified lead to} blob-like structures (cf.~\cite{chattopadhay2018grad,schlemper2019attention}) and, while giving a coarse idea of decision-relevant image areas, still provide only little insight into the decision-making process. 

Model-agnostic methods like LRP~\cite{bach2015pixel}, DeepLIFT~\cite{shrikumar2017learning}, and\linebreak SHAP~\cite{lundberg2017unified}, as demonstrated by Lundberg and Lee, utilize a mechanism called \emph{additive feature attribution} and therefore provide comparable results~\cite{lundberg2017unified}. The basic idea of these methods is to use a backward pass through the network for splitting ''relevance'' starting from the target output, and attributing it to the previous layers with respect to their contributions to the output activation. This procedure is done recursively until the input layer is reached. While employing {\modified a comparable} mechanism, these methods differ in the quality of their generated explanations, with some user studies indicating the most intuitive results for DeepSHAP, a deep learning based approximation of SHAP inspired by DeepLIFT. However, as we will show in Sec.~\ref{Results}, the quality of visualization can be highly dependent on the classifier used, may be poor if the classifier is trained on only few samples, {\modified{}and is prone to adversarial attacks, as pointed out by Kindermans et al.~and Ghorbani et al.~\cite{ghorbani2019interpretation,kindermans2019reliability}. Notably, Apicella et al. presented a general approach for middle-level feature-based explanations based on sparse dictionary methods \cite{ApicellaIPT20} which could successfully be applied to the MNIST and Fashion MNIST datasets, as well as an approach called middle-level feature relevance (MLFR) \cite{apicella2020general}, which generalizes the LRP method from low-level pixel-wise relevance to middle-level features such as super-pixel segmentations for a more intuitive explanation of the relevant input regions. While not being strictly model-agnostic, the approach introduces only few restrictions and is applicable to most deep learning models. Methods like LIME~\cite{ribeiro2016should}, Anchors~\cite{ribeiro2018anchors} and LORE~\cite{guidotti2018local} also fall into this category, but rely on local proxy-models whose adequacy for the task at hand is not guaranteed \cite{apicella2020general,li2020survey}.}

{\modified{}In contrast, image synthesis-based approaches aim for an explanation by creating a synthetic explanation model, and thus can be seen as a special case of model-agnostic approaches. R}ecent {\modified{}work in this domain} attempted to maximize the class output probability of a given class using a gradient ascent optimization on the image input space, such as Activation Maximization (AM) \cite{erhan2009visualizing,simonyan2013deep}. Other work used a more complex approach to achieve this goal, for instance by using a generative adversarial network (GAN)-based activation maximization \cite{nguyen2016synthesizing}. GANs are a method for synthesizing images based on a zero-sum game-like training process \cite{goodfellow2014generative}, and are able to produce images of high quality. AM, however, often results in images of low realism \cite{nguyen2019understanding}, and while this can be improved with GAN-based AM approaches, recent work focused on visualizing the information within the network, rather than the decision-relevant areas for the given image context. {\modified{}Lately, Singla et al.~\cite{singla2019explanation} have proposed a GAN-based approach for visualizing the relevant input regions using an algorithm called progressive exaggeration. Their method aligns well with recent work from Liu et al.~\cite{liu2019generative} which provides GAN-based class prototype and counterfactual images allowing for visual inspection, but extends it by gradual changes and a saliency mechanism. Notably, both approaches were trained on large datasets with 60K-220K images which are rarely available in the medical domain. However, as was shown in previous work, GANs tend to produce only poor results if trained on limited data and may require additional modifications \cite{gurumurthy2017deligan}.}

For clinical acceptance in a medical environment, intuitive visualizations with \emph{realistic images} that are \emph{consistent with the input} are crucial, i.\,e.~the images should reflect the original {\modified{}diagnostic} image as {\modified well} as possible. While having significantly contributed to the scientific State of the Art (SoA) on {\modified{}decision explanation}, current image synthesis-based approaches therefore do not satisfyingly address the task of decision explanation {\modified{}within the domain of clinical decision support}.

\subsection{Outline}
We propose a method which follows the notion of activation maximization and combines it with a cycle-consistent GAN (CycleGAN) to improve the overall realism of the generated decision-explanations. {\modified{}Our method therefore aligns with the work from Singla et al.~\cite{singla2019explanation} and Liu et al.\cite{liu2019generative}, but aims to create more realistic images on smaller datasets. Following the above-mentioned criteria, our method is therefore model-agnostic within the domain of DNNs and can be applied to arbitrary model architectures.} To the best of our knowledge, we are the first to propose a combination of CycleGANs and activation maximization for this purpose. In detail:

\begin{itemize}
    \item We will discuss the steps needed for such an algorithm and explain how each of its parts participates in creating a realistic decision-explanation in Sec.~\ref{GeneralCycleGAN}-\ref{Methods.Visualization}.
    \item In Sec.~\ref{Experiments} we are constructing an experimental setup employing both, a quantitative, as well as a qualitative evaluation technique to evaluate our method in comparison to SoA approaches.
    \item We will present and discuss the results achieved within both experiments in Sec.~\ref{Results}
    \item In Sec.~\ref{Conclusion} an outlook is given, showing up future directions as well as limitations of the study at hand. 
\end{itemize}

\section{Material and Methods}
\label{Methods}
Our proposed method is based on training a cycle-consistent generative adversarial network (CycleGAN~\cite{zhu2017unpaired}, see Sec.~\ref{Methods.CycleGAN}-\ref{Domain transfer}) which is used for activation maximization (Sec.~\ref{Methods.AM}). Within this CycleGAN, each generator is trained to generate images which maximize the activation of one of the output neurons, i.\,e.~class probabilities, of a given two-class classifier for which a decision-explanation should be given. After training (Sec.~\ref{Training_overview}), a difference image of the results of both generators is created, emphasizing decision-relevant regions for arbitrary image inputs (Sec.~\ref{Methods.Visualization}).  

\subsection{Basic Architecture}
\label{GeneralCycleGAN}
\label{Methods.CycleGAN}
\begin{figure}
	\centering
	\includegraphics[width=.7\textwidth]{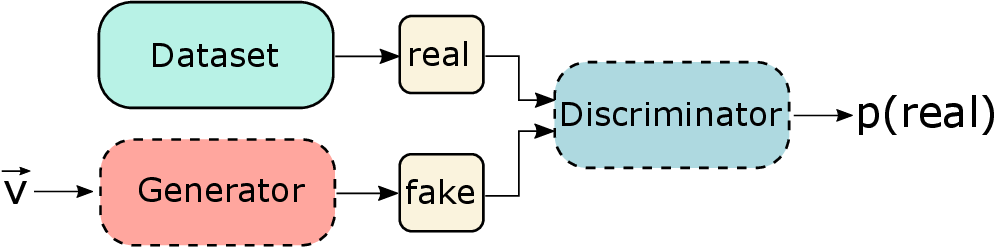}
	\caption{Basic GAN architecture. A generator generates a fake image from a random vector $\Vec{v}$. A discriminator is trained to identify fakes, while the generator is then trained by backpropagating the inverted loss of the discriminator.} 
	\label{Fig:GAN}
\end{figure}

Generally, a CycleGAN consists of a pair of generative adversarial networks (GANs), each consisting of one generator and one discriminator network. In a GAN, a generator network is trained to create synthetic images for the discriminator's image domain. Simultaneously, the discriminator is trained to distinguish real from fake images (see Fig.~\ref{Fig:GAN}).

\begin{figure}
	\centering
	\includegraphics[width=.65\textwidth]{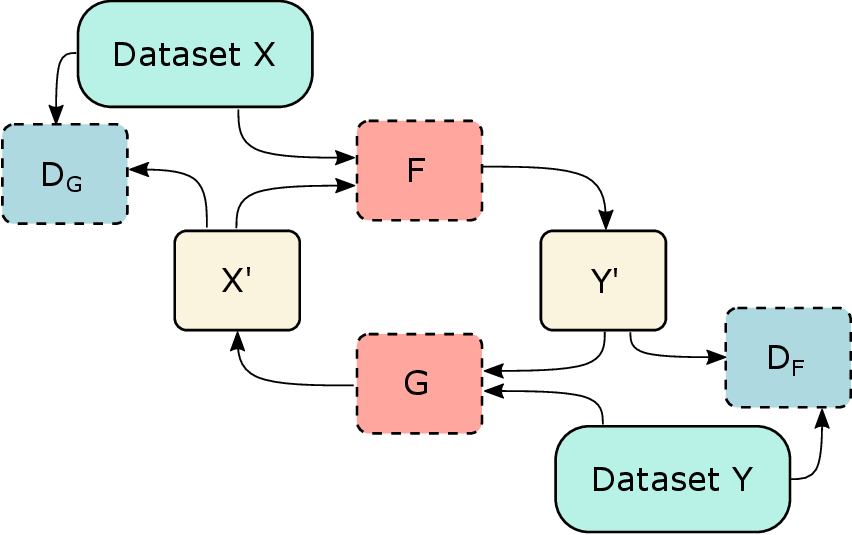}
	\caption{Basic CycleGAN architecture. A pair of GANs $(F,D_F)$ and $(G,D_G)$ is cyclically connected to form a domain transfer between domains $X$ and $Y$ by creating fake images $X',Y'$ using the GAN training concept.} 
	\label{Fig:CycleGAN}
\end{figure}

\begin{figure}
	\centering
	\includegraphics[width=.55\textwidth]{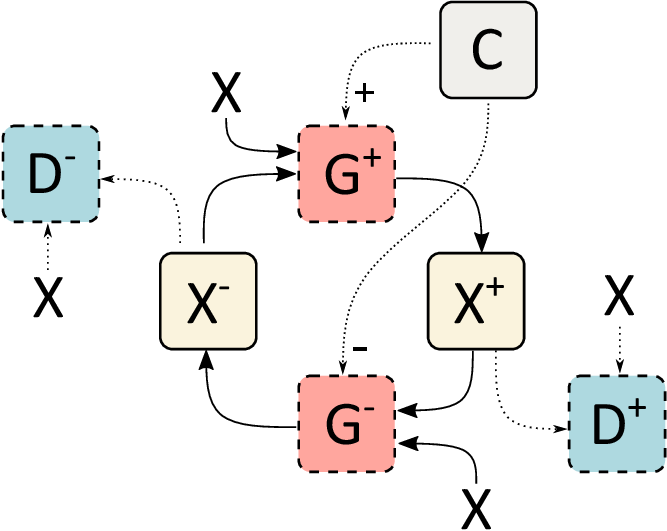}
	\caption{Overview on the general concept of our approach while training with randomly chosen images from image domain $X$. Generators $G^+$ and $G^-$ each maximize one output of the classifier $C${\modified{}, whose decisions are to be explained,} by generating fake images $X^+$ and $X^-$, while discriminators $D^+, D^-$ are used to discriminate real from fake images.} 
	\label{Fig:Concept}
\end{figure}

A CycleGAN employs this process cyclically using image-to-image translations \emph{between} two domains. The discriminator networks are analogously trained to identify \emph{fake images}, i.\,e.~images which originate from the respective other domain, resulting in a zero-sum game like behavior with increasingly growing image realism over the c{\modified{}o}urse of the training \cite{zhu2017unpaired} (see Fig.~\ref{Fig:CycleGAN}). {\modified{}More specifically, a Cycle-GAN is trained to transform images from an image domain $X$ to an image domain $Y$ using a generator $F$ to create a fake samples $Y'$. The same samples are then transformed backwards using a generator $G$ to create fake samples $X'$ in the original image domain. The task of the CycleGAN is to create fake images $Y'$ which are not distinguishable from real images of the original image domain $Y$ as measured by a discriminator $D_F$. Simultaneously, enough information has to be preserved to minimize the error between the original images $X$ and the back-transformed images $X'$. This process is repeated analogously in the other direction, followed by training the discriminators $D_G$ and $D_F$ to recognize fake images, which ultimately leads to continuous improvements in the generated images' quality.}
In our approach we use the same architecture, but in contrast to the original approach train both generators with the \emph{same} image domain and add an additional loss term to each of the generators to maximize one of the output activations {\modified of the classifier whose decisions are to be explained}. An overview of the overall architecture is depicted in Fig.~\ref{Fig:Concept} and will be the basis of the explanations in the following subsections.

\subsection{Cycle Consistency}
\label{CycleConsistency}
CycleGANs are trained to preserve \emph{cycle consistency}, i.\,e.~$G^+(G^-(x)) \approx G^-(G^+(x)) \approx x$ for image inputs $x$ and generators $G^+,G^-$, meaning that a successive mapping through both generators should approximately reconstruct the original input.
To enforce cycle consistency, we use the multiscale structural dissimilarity loss (\emph{MS-DSSIM}) as proposed by Isola et al.~\cite{isola2017image}. Structural dissimilarity considers the image context, luminance, contrast and structure, and ---from a perceptual point of view---is thus preferable to other {\modified{}loss functions} such as L1 or binary cross entropy \cite{wang2003multiscale}. The corresponding loss function reads:
\begin{equation}
\mathcal{L}_{\mathrm{DSSIM}}(x,y) = 1- \frac{(2\mu_x\mu_y+c_1)\cdot(2\sigma_{xy}+c_2)}{2\cdot(\mu_x^2+\mu_y^2+c_1)\cdot(\sigma_x^2+\sigma_y^2+c_2)},
\end{equation}
{\modified{}with stabilizing parameters $c_i=(k_i\cdot L)^2$, $k_1=.01$, $k_2=.03$ and dynamic range $L=1$, in accordance with the original publication}. As we furthermore want to preserve the average intensity, and DSSIM mainly accounts for covariances, we add an L1 term, %
resulting in a cycle consistency loss of
\begin{equation}
\mathcal{L}_{\mathrm{cycle}}(x,x') = \frac{1}{2}(\left|x-x'\right|+\mathcal{L}_{\mathrm{DSSIM}}(x,x'))\\
\end{equation}
This cycle consistency loss is applied to both generator networks $G^+$ and $G^-$ as $\mathcal{L}_{\mathrm{cycle}}(x,G^+(G^-(x)))$ and $\mathcal{L}_{\mathrm{cycle}}(x,G^-(G^+(x)))$, respectively. Additionally, changes to the original image should be kept as small as possible, as only directly decision-relevant regions should be highlighted in the final visualization.
To this end, we add additional similarity losses $\mathcal{L}_{\mathrm{sim.}}(x,G^+(x))$ and $\mathcal{L}_{\mathrm{sim.}}(x,G^-(x))$ using the same loss definition as above with $\mathcal{L}_{\mathrm{sim.}}\equiv\mathcal{L}_{\mathrm{DSSIM}}$.

\subsection{Domain Transfer}
\label{Domain transfer}
We chose the Markovian discriminator approach commonly known as \textit{PatchGAN}~\cite{isola2017image}. In a PatchGAN, the discriminator network's final output is not reduced to a binary (or categorical) variable, but rather preserved as an image-like output, with each output neuron representing the likelihood of its receptive field, i.\,e.~\emph{patch}, to stem from a real image. {\modified This can also be imagined as ``cutting off'' the final layers of a typical neural network for binary classification which narrow down to a single neuron, and instead using the feature maps before the cut-off as the output nodes.} We will call these outputs \textit{likelihood maps}. As shown by Isola et al.~\cite{isola2017image}, the PatchGAN structure prevents the discriminator from focusing on image artifacts, such as the ones created by convolutions at image corners, as otherwise it might not provide useful feedback for the generator anymore. We use multiple intermediate scales, i.\,e.~receptive field sizes, of the same discriminator to further improve the image quality (cf.~\cite{park2019semantic,wang2018high}). 

The PatchGAN structure expects varying difficulties in identifying fake images, depending on the image region under consideration. %
While some regions are easy to assess, e.\,g.~at the image borders (see above), others are rather difficult to detect due to a higher level of image variance. Correspondingly, the loss is more noisy at difficult image regions, and thus provides less stable information for the generator. %
We additionally combat this issue by feeding \emph{two} images at a time to the model, one real and one fake, transforming the task into a categorical decision of \emph{which} of the images is real, thereby creating a more stable feedback and thus facilitate learning---a mechanism comparable to the one used in Siamese GANs~\cite{bashmal2018siamese}. To achieve this, {\modified the} discriminator {\modified extracts feature maps, which we denote likelihood maps,} of {\modified the} two {\modified input} images, one fake and one real, {\modified using shared weights for both}. {\modified Then, a}t each image position, a 2-neuron dense layer with softmax activation is {\modified applied} (locally connected layer in Fig.~\ref{Fig:Discriminator}){\modified{}, the output of which represents a localized discriminator classification of ``realness.'' Please note that this corresponds exactly to the PatchGAN concept described by Isola~et~al.~\cite{isola2017image}, which did not need to be modified for our purposes; the core idea is merely recapitulated for the sake of improved readability.}  %
The {\modified local} discriminators are trained using categorical cross entropy:
\begin{equation}
    \mathcal{L}_{\mathrm{CE}}(y,\hat{y}) = y\cdot\log(\hat{y}) + (1-y)\cdot\log(1-\hat{y})
\end{equation}
where $y$ indicates whether a sample is real or generated and $\hat{y}$ being the discriminator's prediction. The generators are trained by back-propagating the discriminator loss with inverted label information while freezing the discriminator weights. It is noteworthy that none of the generators receive label information \emph{at any time}. However, while both generators are trained \emph{using the same images}, they aim for maximizing opposite class activations (see Fig.~\ref{Fig:Concept}).

\subsection{Activation Maximization}
\label{Methods.AM}
As pointed out in Sec.~\ref{Methods.CycleGAN}, each generator is trained to maximize one of the output activations {\modified of the classifier whose decisions are to be explained} using Activation Maximization (AM). The AM is realized using an additional loss term while training the generators.
{\modified By means of this loss, the classifier to be explained is directly incorporated into the CycleGAN training scheme.} We again use categorical cross entropy $\mathcal{L}_{\mathrm{AM}}(l,\hat{l})=\mathcal{L}_{\mathrm{CE}}(l,\hat{l})$,
where $l$ denotes the label the generator is trained to maximize, e.\,g.~$l=1$ for $G^+$ and $l=0$ for $G^-$, and $\hat{l}$ denotes the classifier's prediction for the current sample. The \emph{actual} label of the sample is not used in this process.

\subsection{Training Overview}
\label{Training_overview}
The generators and discriminators are trained alternately using random pairs of images $(x_a,x_b)$ from the same domain as the {\modified{} input of the classifier to be explained}. First the discriminators {\modified{}$D^+$ and $D^-$}, afterwards {\modified{}the generators $G^+$ and $G^-$} are trained on one batch at a time. 
Combining the loss functions from above, the overall generator loss for $G^+$ reads:
\begin{equation}
\begin{aligned}
\mathcal{L}_{\mathrm{gen}}(x_a,x_b,l,G^+,G^-,D^+,C)=&\mathcal{L}_{\mathrm{cycle}}(x_a,G^-(G^+(x_a))) &+&\hspace{.5em}\mathcal{L}_{\mathrm{sim.}}(x_a,G^+(x_a))\hspace{.5em}+\\ & \mathcal{L}_{\mathrm{CE}}(D^+(G^+(x_a),x_b),0) &+&\hspace{.5em}\mathcal{L}_{\mathrm{AM}}(l,C(x_a))
\end{aligned}
\end{equation}
with $x_a,x_b$ being images in the {\modified{}image domain $X$ of the classifier $C$ to be explained}, the label $l$ which should be maximized by the generator $G^+$, its respective discriminator $D^+$, and $G^-$ being the opposite generator. The loss function is analogously applied to generator $G^-$ with inverted $+$ and $-$. The procedure is repeated until the overall loss function of both generators converges. Using this loss definition, the CycleGAN is trained to
\begin{enumerate}
	\item maximize the output probability of the class assigned to each of the generators (\textbf{AM loss}) with only small changes (\textbf{similarity loss}), 
	\item generate images indistinguishable from real images (\textbf{discriminator loss}),
	\item be cycle-consistent (\textbf{cycle-consistency loss})
\end{enumerate}

\subsection{Relevancy Visualization}
\label{Methods.Visualization}
Based on the loss function above, the generators are trained to provide two slightly modified versions $G^+(x),G^-(x)$ of the original images $x$ for which the decision of the classifier should be explained, with each of them maximizing one of the classifier's output activations. As the generators are trained to introduce as few changes as possible (\emph{similarity loss}), the modified regions are expected to be immediately decision-relevant, while the image is kept consistent to the original input. Assuming that the generators $G^+$ and $G^-$ are trained on the same data set and realize the transitions $C(G^+(x))\rightarrow1$ and $C(G^-(x))\rightarrow0$, respectively, differences between $x$, $G^+(x)$, and $G^-(x)$ can be attributed to semantic differences with respect to the classifier's decision strategy. {\modified{}Let us define the difference map per class as:}
\begin{equation}
\Delta^+ = \Delta(G^+(x),x),\hspace{3em} \Delta^- = \Delta(G^-(x),x)
\end{equation}
{\modified{}with $\Delta$ being a local difference metric. While the method itself is not restricted to any specific metric, we suggest using pixel-wise absolute difference or a patch-wise structural dissimilarity. While pixel-wise absolute difference will work well in most cases, this choice might be dataset specific, depending on whether structural or intensity changes shall be emphasized.} We {\modified{}can now} define overall relevance $R$ as:
\begin{equation}
\begin{aligned}
R &= \Delta(G^-(x),x) - \Delta(G^+(x),x) \\
&=\Delta^- - \Delta^+ \\
\end{aligned}
\end{equation}
{\modified{}This definition reflects that image changes which have to be introduced to maximize one class are an indicator of the relevance of that region for the respective other class within the given context.} Assuming a successful training of $G^+$ and $G^-$ according to the above-mentioned targets, $R$ has the following properties: it shows low absolute magnitudes in areas for which a modification would not result in {\modified{}any} activation modification for the classifier {\modified{}to be explained} {\modified{}or would have to be modified equally for both classes}, and high magnitudes in areas either only relevant for $G^+$, or $G^-$, as they would result in an output modification for the classifier. Therefore, $R$ is an immediate indicator of decision relevance.

\section{Experimental}
\label{Experiments}
{\modified{}To show the adequacy of our algorithm for different architectures and scenarios, we conducted studies on three different datasets which are all publicly available. For the evaluation, we chose a two-step experimental setup.}
In each study, we first train a classification network for a binary decision task. After validating its ability to separate positive and negative class samples, our method for decision explanation is trained on the created classifier as described in Sec.~\ref{GeneralCycleGAN}-\ref{Methods.Visualization}. %
For ensuring a successful training of the CycleGAN, a quantitative evaluation is done, analyzing whether the generated samples are indeed modifying the class output probabilities of the classifier {\modified{}to be explained} in the expected way.
In a second step, a double-blind user study is employed to compare our method to various SoA approaches for decision explanation. The generated results are comprehensively evaluated and statistically tested.

{\modified{}
\subsection{Datasets}
\paragraph{LIDC-IDRI} For the the first study, we} used the well-known LIDC-IDRI computed tomography dataset for lung lesion malignancy estimation~\cite{armato2015data,armato2011lung,clark2013cancer}. The data was preprocessed in accordance with prior work by Nibali et~al.~\cite{nibali2017pulmonary}. Lesions were extracted as axial slices at 64\,x\,64 pixels using bilinear interpolation and represented a window of 45\,x\,45\,mm in world coordinates around the lesion's midpoint. Only lesions with at least three radiological annotations were used, assuming a lesion with a median malignancy rating above three to be malignant and below benign. As in Nibali et~al.~\cite{nibali2017pulmonary}, samples with a borderline malignancy rating of three were discarded to enforce separability. The resulting data set contained $772$ samples with 348 positive and 424 negative samples. 70\,\% of the data were used for training and validation (236/301), 30\,\% as test set (112/123).
{\modified{}
\paragraph{BreastMNIST} \label{Datasets_BreastMNIST}Secondly, we used the BreastMNIST dataset from Yang et al.~\cite{yang2020medmnist}. BreastMNIST is part of MedMNIST,  a standardized, diverse and lightweight data collection for model evaluation in a medical setting. It consists of 780 breast ultrasound images with or without signs of malignant lesions. In this study, the dataset is used to show the applicability of our approach to another medical imaging modality. The original training, validation and test splits were used, resulting in 624 samples for training and validation (456 positive/168 negative), and 156 samples for testing (114/42). Each sample has a resolution of 28\,x\,28 pixels which was nearest neighbor-padded to 32\,x\,32 pixels for an easier handling in the model architecture.}
{\modified{}
\paragraph{CIFAR-10} Thirdly, we employed the CIFAR-10 dataset from Krizhevsky et al.~\cite{krizhevsky2009learning} consisting of 50,000 images for training and validation, and 10,000 images for testing with a resolution of 32\,x\,32 pixels. The dataset consist of 10 different classes, where each class is represented by 5,000/1,000 images for training and validation, or testing, respectively. As our approach is explicitly tailored to binary classification tasks, we created two subsets based on the most difficult 1-vs-1 class decisions which turned out to be cats-vs.-dogs and cars-vs.-trucks, as shown in Sec.~\ref{Quantitative_Results}.}

\subsection{Architecture and Training}
\begin{figure}
\centering
\includegraphics[width=\textwidth]{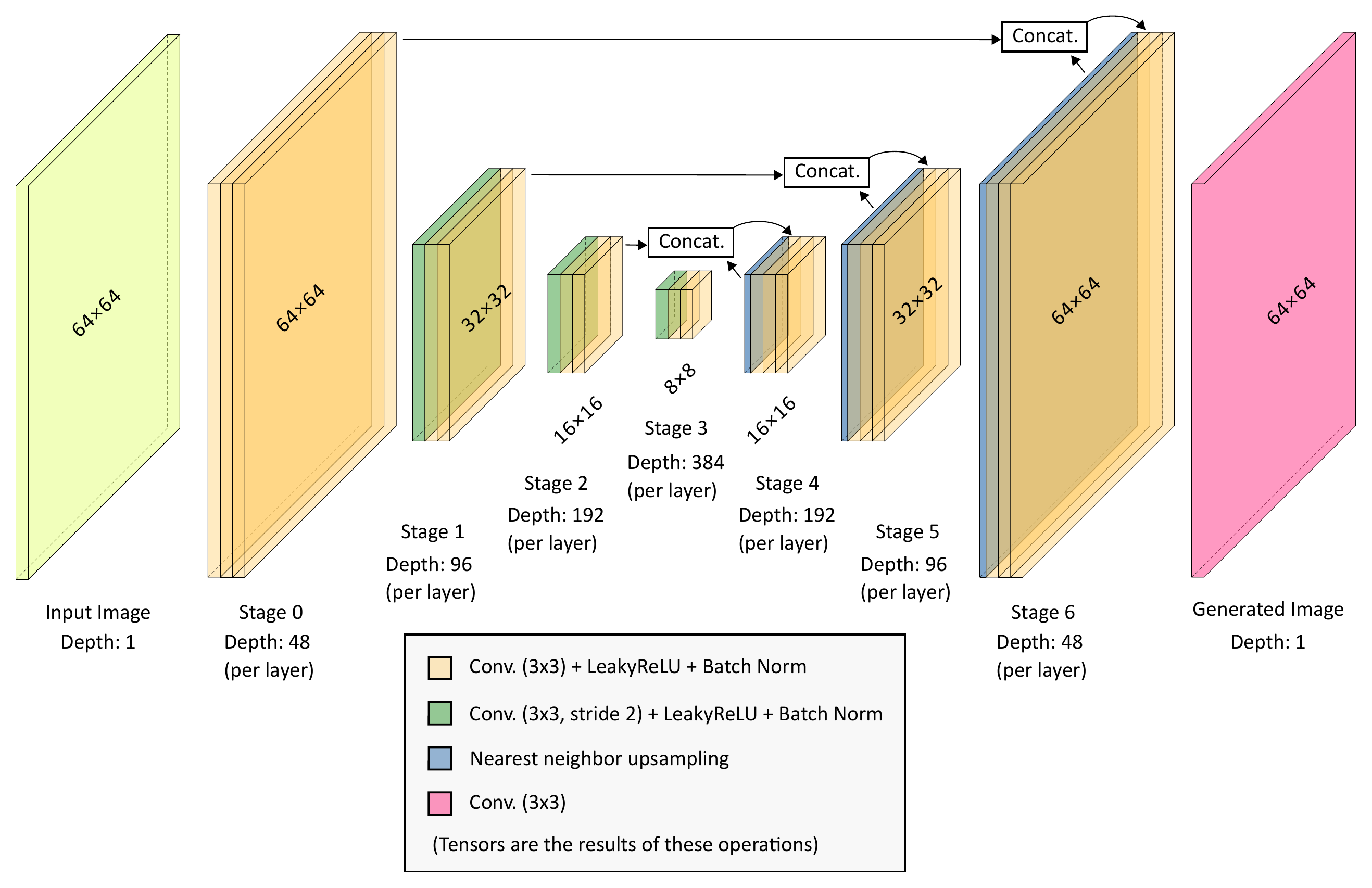}
\caption{Overview on the generator network architecture used for domain transfer in the CycleGAN. The architecture is based on U-Net \cite{ronneberger2015u} using LeakyReLU and convolutions {\modified with strides}. We applied 3 + 1 stages of 3 convolutional layers per stage (beige), each followed by LeakyReLU activation and batch normalization. Upsampling (blue) is done using nearest neighbor interpolation.}
\label{Fig:Generator}
\end{figure}

\begin{figure}
\centering
\includegraphics[width=\textwidth]{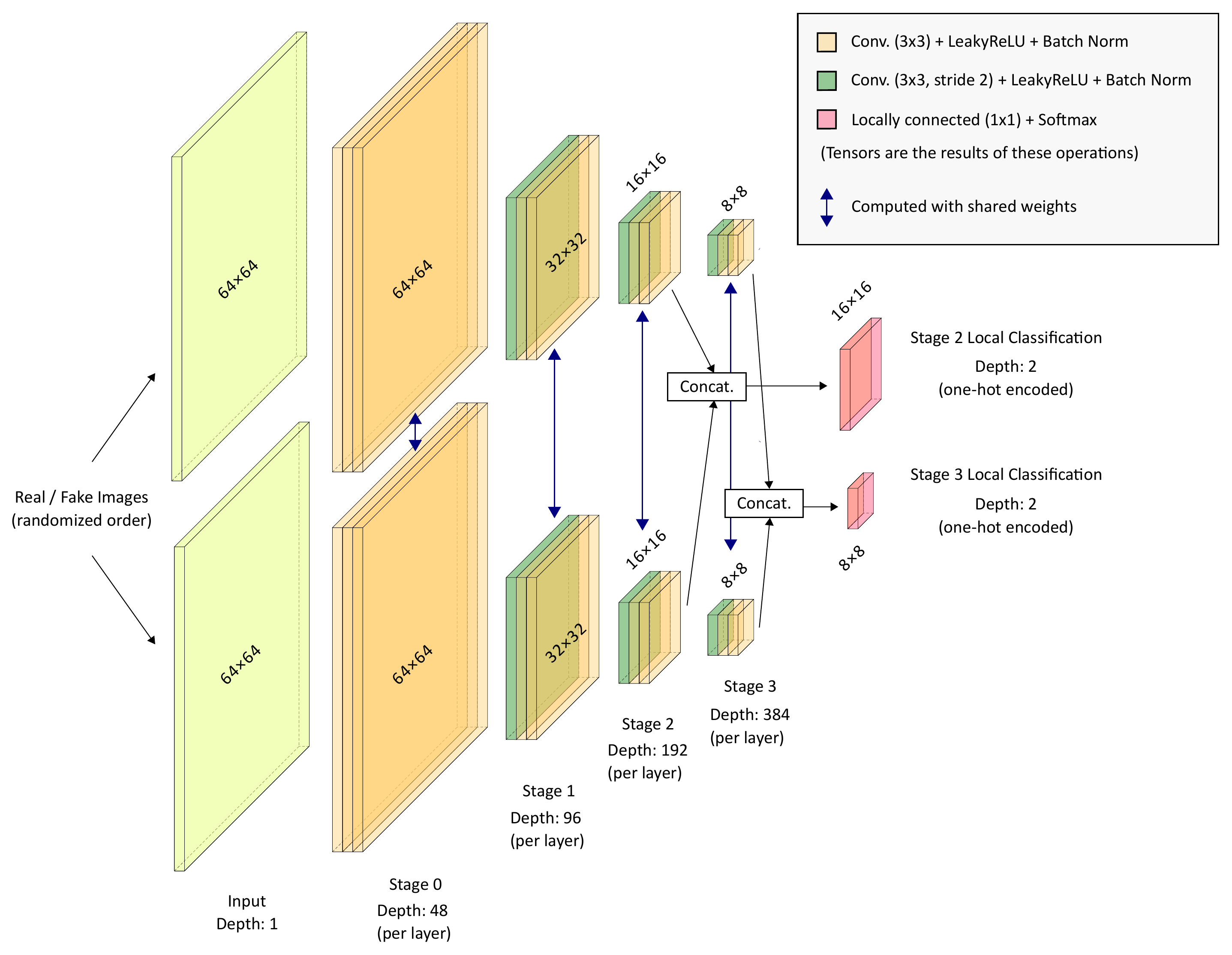}
\caption{Overview on the discriminator network architecture. The architecture resembles the generator architecture from Fig.~\ref{Fig:Generator}, but is cut off after stages 2 and 3 for multiscale patch classification (cf.~\cite{park2019semantic,wang2018high}). Each output neuron represents a receptive field {\modified in} the input space (PatchGAN). Feature extraction is done on two images simultaneously, one real and one fake with randomized order, using weight sharing. The final classification is done by concatenating the features and appending a 2-neuron softmax output for each pair of {\modified nodes} (see Sec.~\ref{Domain transfer}).}
\label{Fig:Discriminator}
\end{figure}

{\modified{}
\paragraph{LIDC-IDRI}~For the lung lesion dataset, a} ConvNet was used for malignancy classification, built of four convolutional blocks using 3x3 convolutions with a stride of 1 in the first and 2 for the subsequent layers, followed by batch normalization and ReLU activation. In order, the blocks had 32, 64, 128 and 256 kernels. After flattening, a softmax layer was appended. The classifier was trained until convergence using weighted categorical cross entropy. Afterwards, our method for decision explanation was trained. The CycleGAN's generator networks were based on U-Net \cite{ronneberger2015u}, using a depth of 3, with 3 convolutions per stage. The convolutional layers had 48, 96, 192 and 384 kernels at stages 0, 1, 2 and 3, respectively. The discriminators were based on the same architecture, but clipped to the outputs of stages 2 and 3 (cf.~PatchGAN, see Sec.~\ref{Domain transfer}). The architecture of the generator and discriminator networks are depicted in Figs.~\ref{Fig:Generator} and \ref{Fig:Discriminator}, respectively. {\modified Absolute intensity difference was used as the difference measure for relevancy visualization.}

{\modified{}
\paragraph{BreastMNIST}~To show the applicability of our algorithm to other classifier architectures, we used the ResNet Keras reference implementation from Chollet \cite{chollet2015keras} for the BreastMNIST dataset. As we expected textural rather than intensity changes, we used the structural dissimilarity as the difference metric. No other modifications were applied. Similarly, the generator and discriminator architectures were left unchanged.
\paragraph{CIFAR-10} After training the prior architectures from scratch, we wanted to demonstrate the value of our method for highly complex, pretrained architectures, where we expect the compared reference algorithms to have their specific strengths. Therefore we used the EfficientNet-B0 implementation from Yakubovskiy \cite{yakubovskiy2020efficientnet} based on the work from Tan et al.~\cite{tan2019efficientnet} and fine-tuned it on the CIFAR-10 dataset. To account for the vastly larger amount of data in comparison to our previous scenario, we increased the number of kernels per layer for both the generator as well as the discriminator models to 256 filters in the first layer, but followed the same scheme as before. The maximum number of kernels per layer was clipped to 512 to account for the limited GPU memory, resulting in 256, 512, 512, 512 kernels at stages 0, 1, 2 and 3, respectively. Similarly, structural dissimilarity was used as a relevancy measure. The architecture itself was identical to the first and second study.}

\subsection{Quantitative Evaluation}
For the quantitative evaluation, we first evaluate{\modified{}d} whether the trained classifier is adequate for the task at hand. Therefore we {\modified{}assessed} the classifier's performance for {\modified{}the classification task} on the test set using various classification metrics, such as accuracy, sensitivity, specificity, positive and negative predictive value, informedness, markedness, Matthews correlation coefficient and area under the curve (AUC). After ensuring the classifier is applicable, we evaluated whether our proposed CycleGAN network is capable of a successful domain transfer by evaluating the classifier's output class probabilities on the original data, as well as the modified versions, i.\,e.~the outputs of $G^+$ and $G^-$, and thus whether a change in the indicated regions is associated with an actual modification of the class output probabilities in the expected way.

\subsection{User Study}
\label{Experimental.UserStudy}
\begin{figure}
	\centering
	\includegraphics[width=\textwidth]{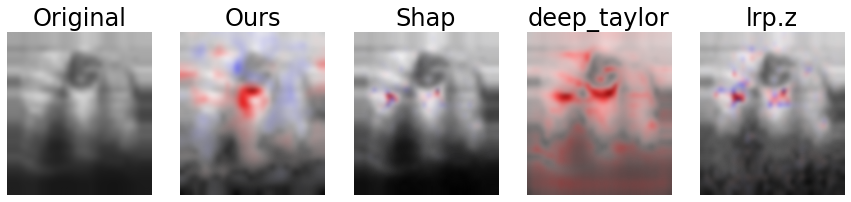}\\
	\includegraphics[width=\textwidth]{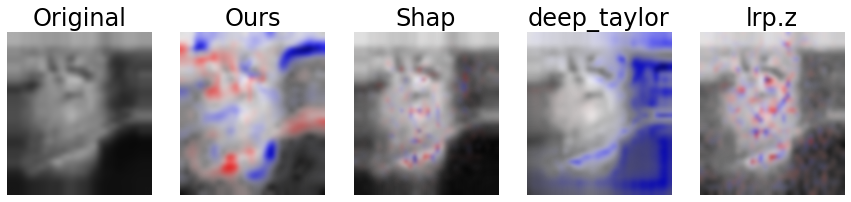}\\
	\vspace{2em}
	\includegraphics[width=\textwidth]{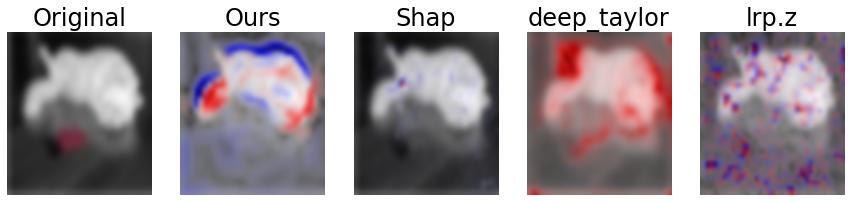}\\
	\includegraphics[width=\textwidth]{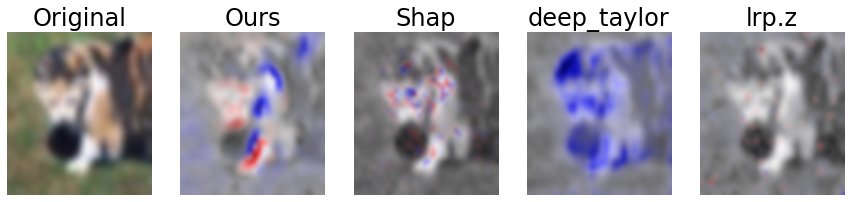}\\
	\vspace{1em}
    \caption{\modified{}Example visualizations of misclassified samples on the BreastMNIST (top) and CIFAR-10 cats-vs.-dogs (bottom) datasets for false positives (first, third) and false negatives (second, fourth). Notably, our algorithm highlights regions with respect to their contribution while DeepTaylor tends to mark all relevant regions with the incorrectly predicted label. To avoid mixing up classification and visualization errors, misclassified samples were excluded from the user study.}
	\label{Fig:Misclassified}
\end{figure}

For the qualitative evaluation, we asked $N=8$ medical engineers with multiple years of experience in the field of medical algorithm development (average: 6,5 years) to evaluate the results {\modified{}on the LIDC-IDRI dataset} in terms of a) \emph{intuitive validity} of the visualization (,,Does it look reasonable at a glance?''), b) \emph{semantic meaningfulness} in the context of {\modified{}the respective classification task} (,,Does it make sense?''), and c) overall \emph{image quality} (,,Does it look good?''). {\modified{}Due to the coronavirus pandemic, the studies on the BreastMNIST and CIFAR-10 datasets were conducted using an online-survey which was similarly organized.
For the CIFAR-10 dataset, $N=9$ participants for the cats-vs.-dogs and $N=6$ participants for the trucks-vs.-cats scenario completed the questionnaire. The BreastMNIST survey was completed by 12 participants from which one participant was excluded due to an incomplete survey, leaving $N=11$ participants for the final evaluation. Similar to the LIDC-IDRI evaluation, the participation in the BreastMNIST study was restricted to participants with significant experience in medical image processing and/or assessment.} 
For each criterion we analyzed the inter-observer reliability using pairwise Pearson product-moment correlation $\rho$ over all participants with questionnaires Q as:
\begin{equation}
    \rho=\frac{1}{\binom{N}{2}}\sum_{i=1}^{N-1}\sum_{j=i+1}^{N} r(Q_i,Q_j),
\end{equation}
We double-blindly evaluated our method against DeepSHAP~\cite{lundberg2017unified} using the reference implementation from Lundberg~et~al., as well as DeepTaylor~\cite{montavon2017explaining} and LRP~\cite{bach2015pixel} using the iNNvestigate libary from Alber~et~al.~\cite{alber2019innvestigate}. Each participant received images of 24 {\modified{}samples (12 positive, 12 negative)} with the classifier's decisions being visualized by each of the methods, depicted as a colored overlay over the original lesion image, {\modified{} resulting in a total of $192$ answers per criterion, algorithm and task, $288$ samples per questionnaire, or $9{,}792$ answers in total}. 
The image order was randomized for each question by a computer program to avoid an identification of the decision-explanation algorithm which generated the image. {\modified{}For the LIDC-IDRI dataset, two questionnaire variants (A and B) have been generated to find out whether the presented order has an influence on the perceived quality \cite{krosnick2018questionnaire}. After we ensured that this was not the case (see Sec.~\ref{Results_UserStudy}), we dropped this step for the BreastMNIST and CIFAR-10 datasets}. The images were randomly drawn from the correctly classified test set samples to avoid mixing up classification and visualization errors {\modified{}(cf.~Fig.~\ref{Fig:Misclassified}).} The participants were asked to assign an integral score from -4 (low) to 4 (high) for each criterion to each image. Afterwards, the scores were collected and de-randomized by the computer program again. All responses were adjusted by z-normalizing over all answers for the same {\modified{}image} and criterion. The results of the different decision explanation methods were compared against each other with respect to their average adjusted scores as well as their average rank scores and statistically tested using a two-tailed $t$-test with $t(N-2)$. Finally, after verifying its applicability using the Kaiser-Meyer-Olkin measure \cite{kaiser1970second,dziuban1974correlation}, a principal component analysis (\emph{PCA}) was employed to analyze whether the \emph{preferability} of an image can be described by a general factor $\Phi$ that covers most of the variance.

\section{Results}
\label{Results}

\subsection{Quantitative Results}
\label{Quantitative_Results}

\begin{table}[]
\centering
\begin{tabular}{ccc}
\toprule
Metric & LIDC-IDRI & BreastMNIST \\
\midrule
Accuracy     & .809 {[}.757,.860{]} & .802 {[}.737,.859{]} \\
F1           & .801 {[}.737,.856{]} & .871 {[}.824,.913{]} \\
Sensitivity  & .813 {[}.737,.884{]} & .921 {[}.869,.966{]} \\
Specificity  & .805 {[}.729,.866{]} & .477 {[}.326,.629{]} \\
PPV          & .790 {[}.715,.861{]} & .827 {[}.760,.890{]} \\
NPV          & .826 {[}.757,.894{]} & .690 {[}.515,.851{]} \\
Informedness & .618 {[}.514,.719{]} & .398 {[}.239,.557{]} \\
Markedness   & .616 {[}.511,.718{]} & .517 {[}.329,.691{]} \\
MCC          & .617 {[}.511,.718{]} & .453 {[}.285,.607{]} \\
AUC          & .809 {[}.757,.860{]} & .822 {[}.740,.895{]} \\
\bottomrule
\end{tabular}
\caption{\modified{}Test set performance of the used classifiers for malignancy classification on the LIDC-IDRI and BreastMNIST datasets with 95\,\% CI.}
\label{Tab:Classifier_ResultsLIDCMNIST}
\end{table}

\begin{table}[]
\centering
\begin{tabular}{cccc}
\toprule
Metric & CIFAR-10 & Cats-vs.-Dogs & Trucks-vs.-Cars \\
\midrule
Accuracy     & .960 {[}.959,.963{]} & .937 {[}.926,.947{]} & .980 {[}.974,.985{]} \\
F1           &                      & .937 {[}.926,.947{]} & .979 {[}.974,.985{]} \\
Sensitivity  &                      & .943 {[}.929,.957{]} & .970 {[}.959,.980{]} \\
Specificity  &                      & .931 {[}.914,.946{]} & .988 {[}.983,.994{]} \\
PPV          &                      & .932 {[}.917,.947{]} & .988 {[}.982,.994{]} \\
NPV          &                      & .942 {[}.927,.957{]} & .971 {[}.960,.980{]} \\
Informedness &                      & .874 {[}.853,.895{]} & .959 {[}.948,.970{]} \\
Markedness   &                      & .874 {[}.852,.894{]} & .960 {[}.948,.970{]} \\
MCC          & .956 {[}.954,.959{]} & .874 {[}.853,.894{]} & .960 {[}.948,.970{]} \\
AUC          & .978 {[}.977,.979{]} & .987 {[}.984,.991{]} & .997 {[}.995,.999{]} \\
\bottomrule
\end{tabular}
\caption{\modified{}Test set performance of the used EfficientNet-B0 model on the whole CIFAR-10 and the cats-vs.-dogs and trucks-vs.-cars subsets with 95\,\% CI. Two-class metrics were left out for the 10-class model. For the two-class models the first named was encoded as the positive class.}
\label{Tab:Classifier_ResultsCIFAR}
\end{table}

\paragraph{Classifier Performance}
The employed {\modified{}LIDC-IDRI classifier} achieved a sensitivity of {\modified{}.813 95\% CI [.737,.884] and a specificity of .805 [.729,866] with an area under the curve (AUC) of .809 [.757, .860].}
{\modified{}The respective values for the BreastMNIST dataset were .921 {[}.869,.966{]}/.477 {[}.326,.629{]}/.822 {[}.740,.895{]} for sensitivity, specificity and AUC. There was a clear difference between the sensitivity and the specificity of the BreastMNIST classifier. Similarly we needed to train multiple classifiers to receive explanations from the DeepTaylor approach. Both of this can likely be attributed to the dataset's strong label imbalance (cf. Sec.~\ref{Datasets_BreastMNIST}). The CIFAR-10 classifier had an overall class accuracy of .960 {[}.959,.963{]} over all 10 classes. The accuracies for the cats-vs.-dogs and trucks-vs.-cars classifiers were .937 {[}.926,.947{]} and .980 {[}.974,.985{]}, with AUCs of .987 {[}.984,.991{]} and .997 {[}.995,.999{]}. Notably, all of the aforementioned classifiers should constitute} a reasonable baseline for decision explanation. More detailed performance results can be found in Tab.~\ref{Tab:Classifier_ResultsLIDCMNIST} and \ref{Tab:Classifier_ResultsCIFAR}.

\begin{figure}
	\centering
	\includegraphics[width=.75\textwidth]{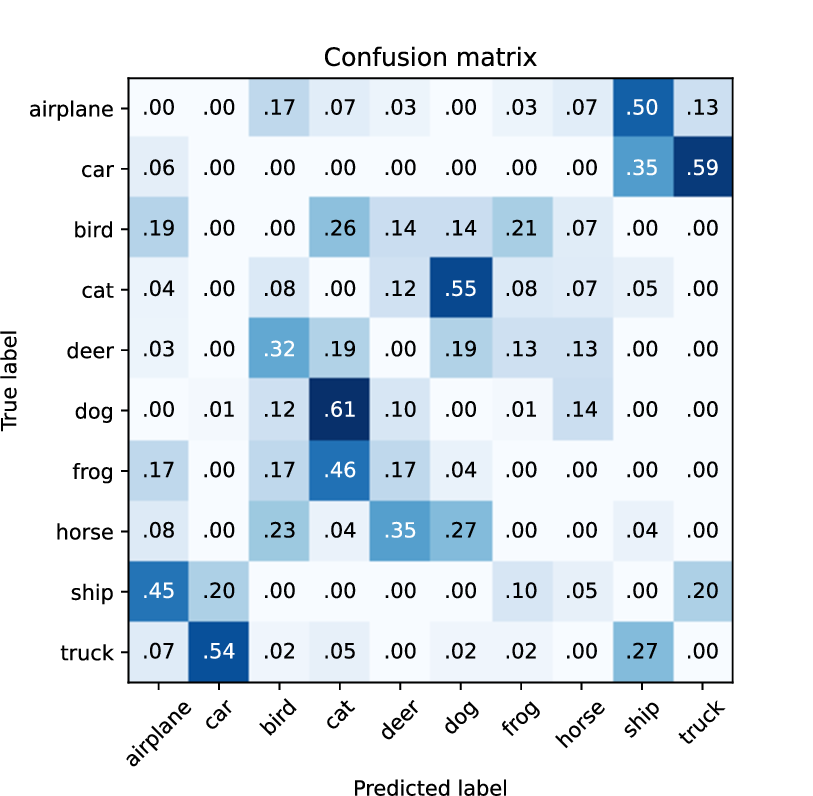}
	\caption{\modified{}The relative proportions of errors between class pairs on the CIFAR-10 dataset (main diagonal eliminated), with cats-vs.-dogs and trucks-vs.-cars accounting for over 50\,\% of the errors for each of the classes.}
	\label{Fig:CIFAR_Confusion}
\end{figure}

\paragraph{Domain transfer}
{\modified{}
\paragraph{LIDC-IDRI}~}
The classifier's average malignancy estimates{\modified{}, i.\,e.~the arithmetic means of the classifier outputs for the \emph{malignant} class,} on the test data were .470, 95\,\% CI [.418,.522] before modification, .289 [.250,.330] after negative transfer, and .820 [.788, .848] after positive transfer, i.\,e.~$x$, $G^-(x)$ and $G^+(x)$. The pairwise differences between the transferred and the original domains were each highly significant with $p\ll 10^{-5}$ (two-tailed $t$-test, $t(233)=17.9$ for $x/G^+(x)$, $14.5$ for $x/G^-(x)$), implying a substantial modification of the classifier's class output probabilities using the proposed CycleGAN architecture.
{\modified{}

\paragraph{BreastMNIST} Similar values could be observed for the BreastMNIST dataset with an average malignancy rating of .656, 95\,\% CI [.604,.707] before, .337 [.294,.383] after negative and .871 [.835,.904] after positive transfer. All differences were highly significant with $p\ll 10^{-5}$ ($t(154)=10.2/13.4$).

\paragraph{CIFAR-10} The values for the CIFAR-10 cats-vs.-dogs and trucks-vs.-cars transfers were .504 [.482,.525]/.491 [.469,.513] before transfer, .152 [.138,.165]/.179 [.165,.194] after negative, and .920 [.911/.930]/.863 [.850,.876] after positive transfer. Again, all differences were highly significant ($t(998)\ge 33.3$).

Notably, for each dataset and architecture the generators $G^-(x)$ and $G^+(x)$ were able to generate images which are recognized significantly more likely as the respective class than the original image.
}

\subsection{User Study}
\label{Results_UserStudy}
\begin{figure}[]
	\centering
	\includegraphics[width=.45\textwidth]{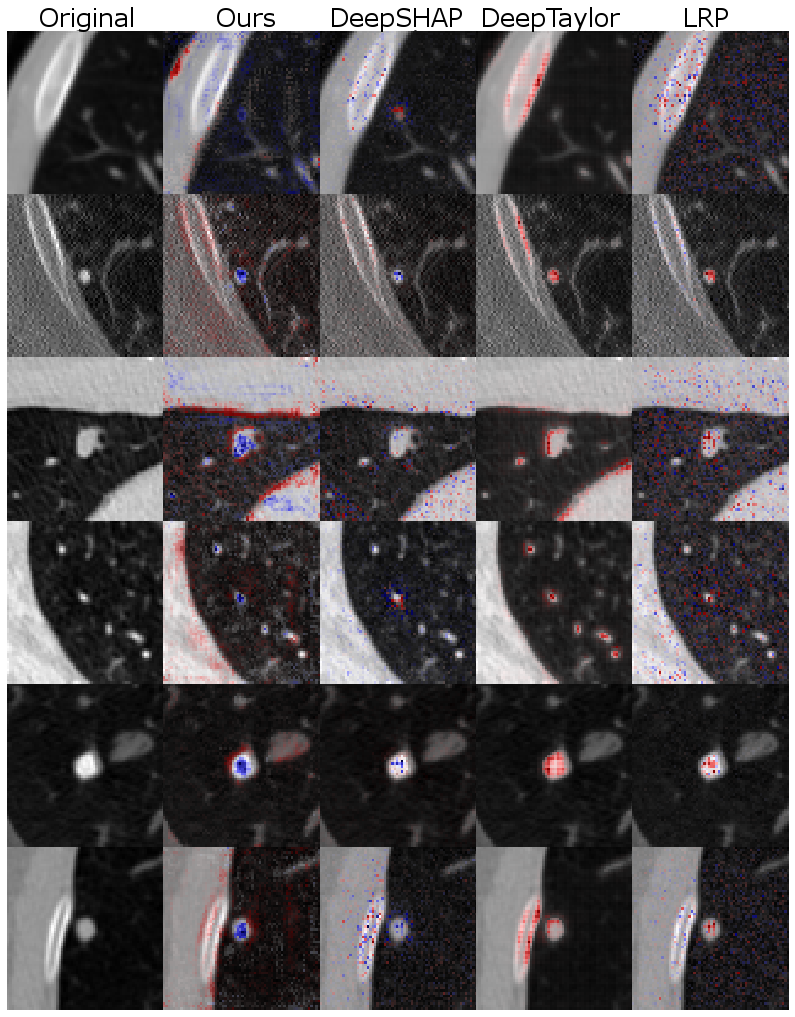}\hspace{1em}
	\includegraphics[width=.45\textwidth]{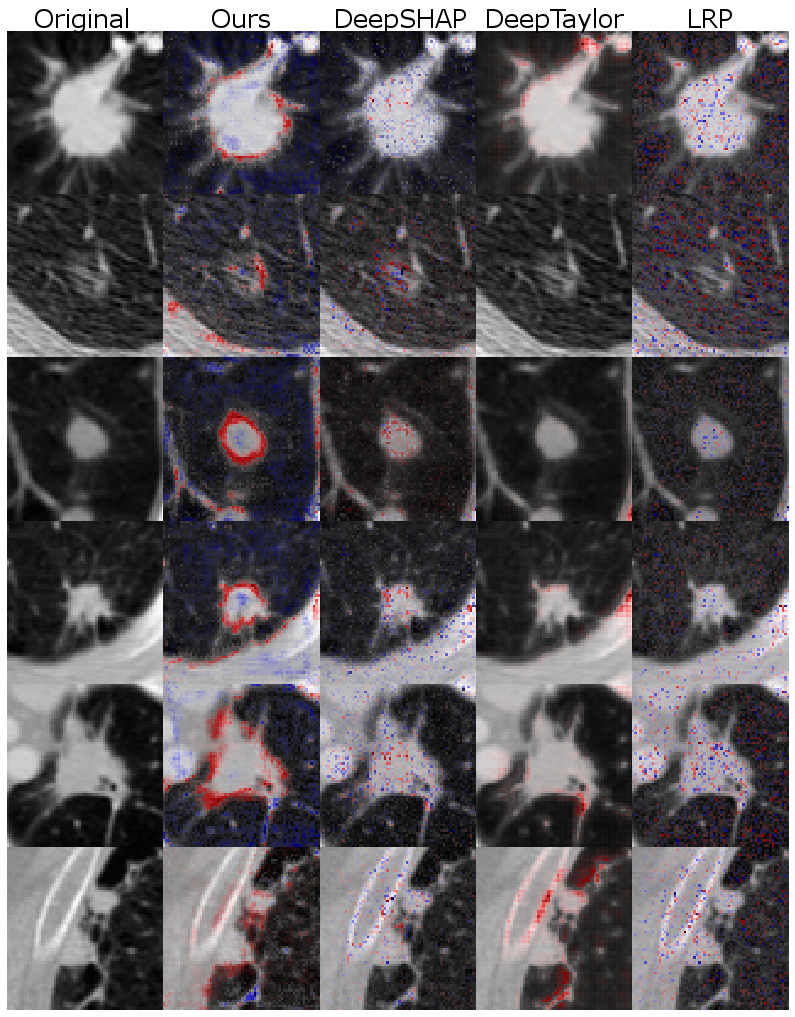}
	\caption{Example images for benign (left) and malignant (right) lung lesions with blue/red indicating benignity/malignancy-indicating regions {\modified{}on the LIDC-IDRI dataset}.
	}
	\label{Fig:Visualization}
\end{figure}

\begin{figure}[]
	\centering
    \includegraphics[width=.45\textwidth]{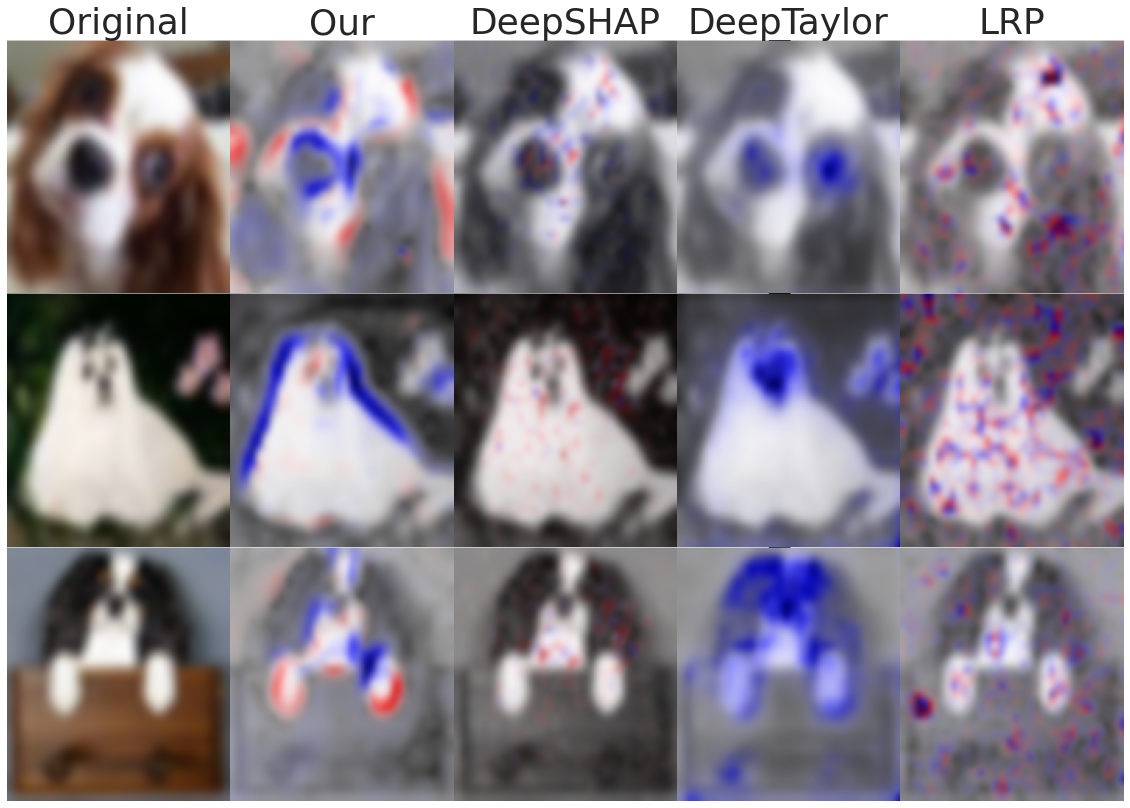}\hspace{1em}
	\includegraphics[width=.45\textwidth]{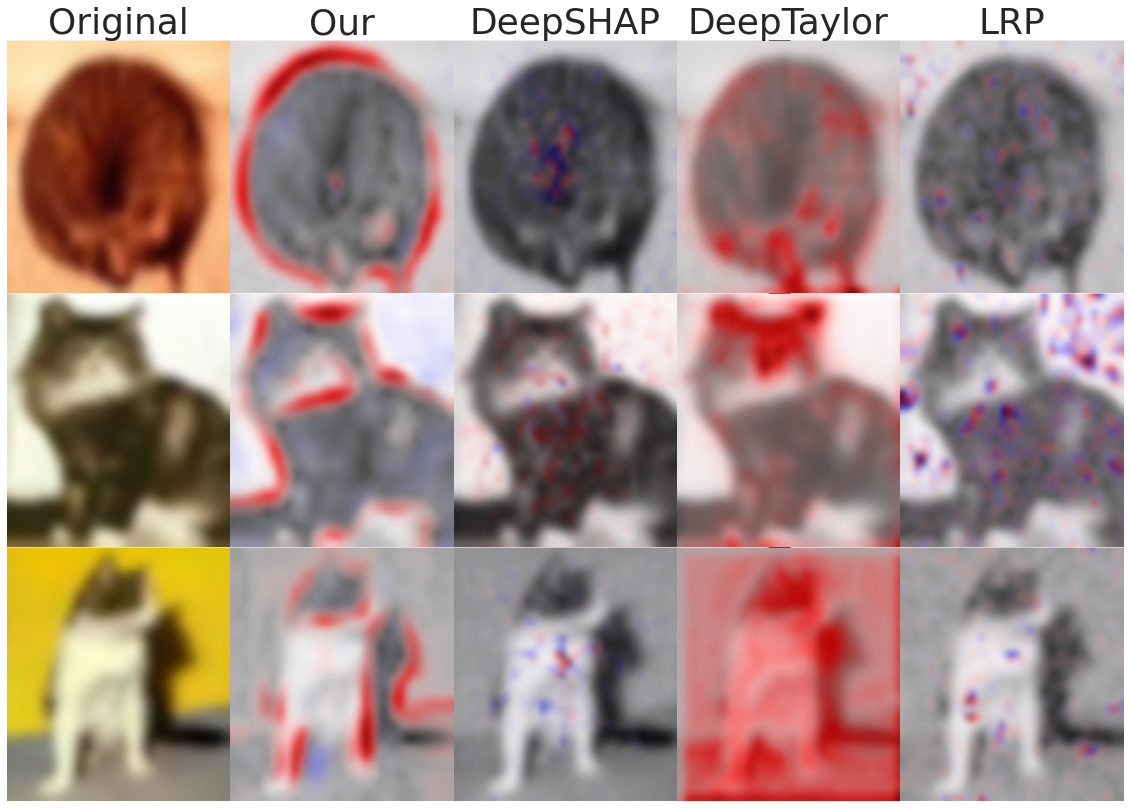}\\\vspace{1em}
	\includegraphics[width=.45\textwidth]{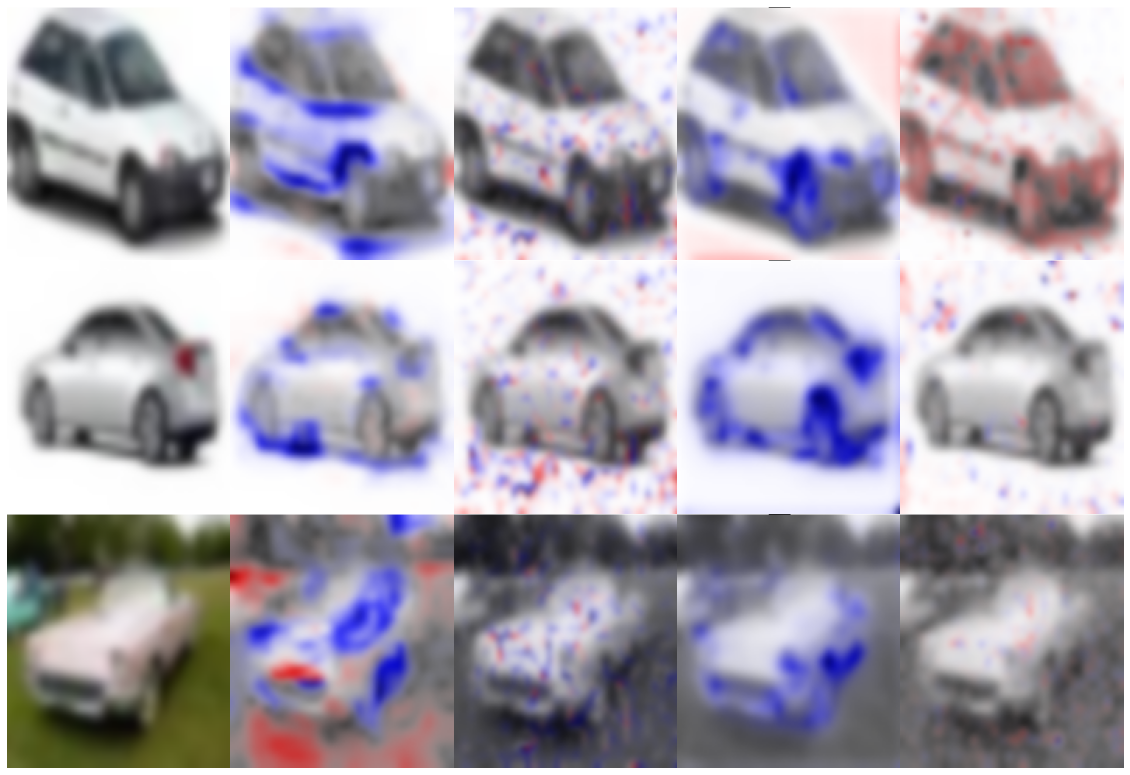}\hspace{1em}
	\includegraphics[width=.45\textwidth]{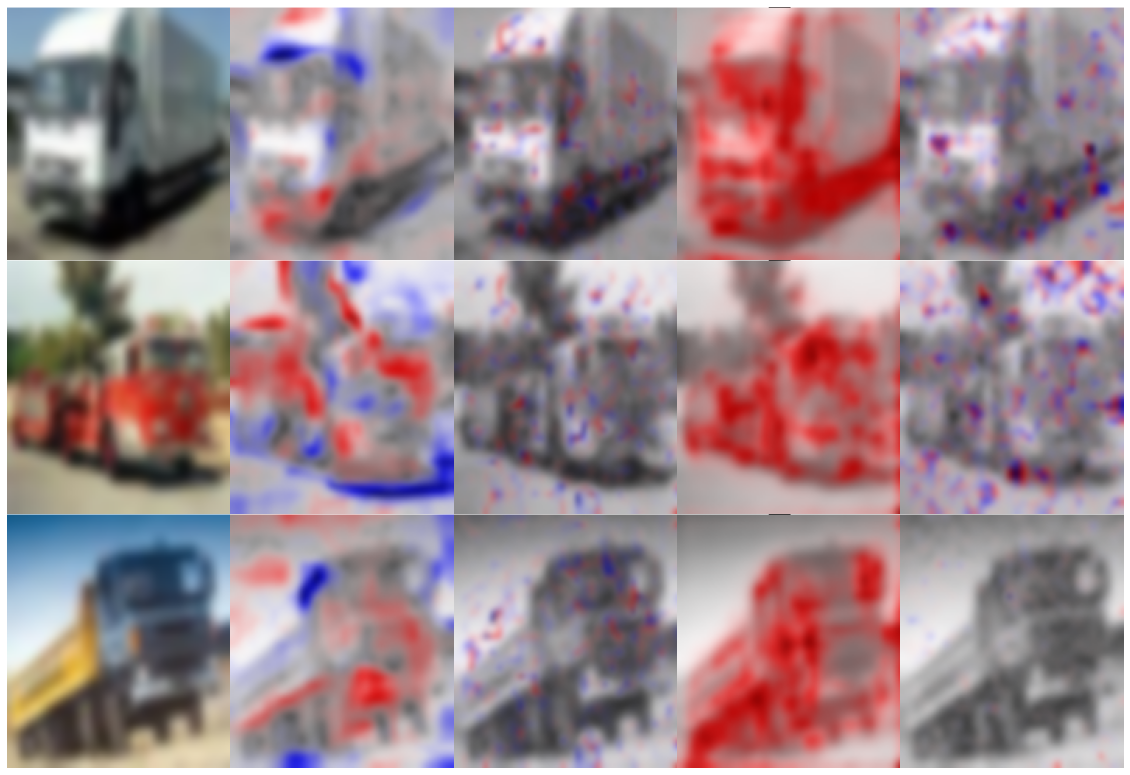}\\
	
	\caption{\modified{}Example images on the CIFAR-10 dataset (top: cats-vs.-dogs; bottom: cars-vs.-trucks). Dog- and car-likeness was encoded blue, cat- and truck-likeness was encoded red.
	}
	\label{Fig:Visualization_CIFAR}
\end{figure}

\begin{figure}[]
	\centering
    \includegraphics[width=.45\textwidth]{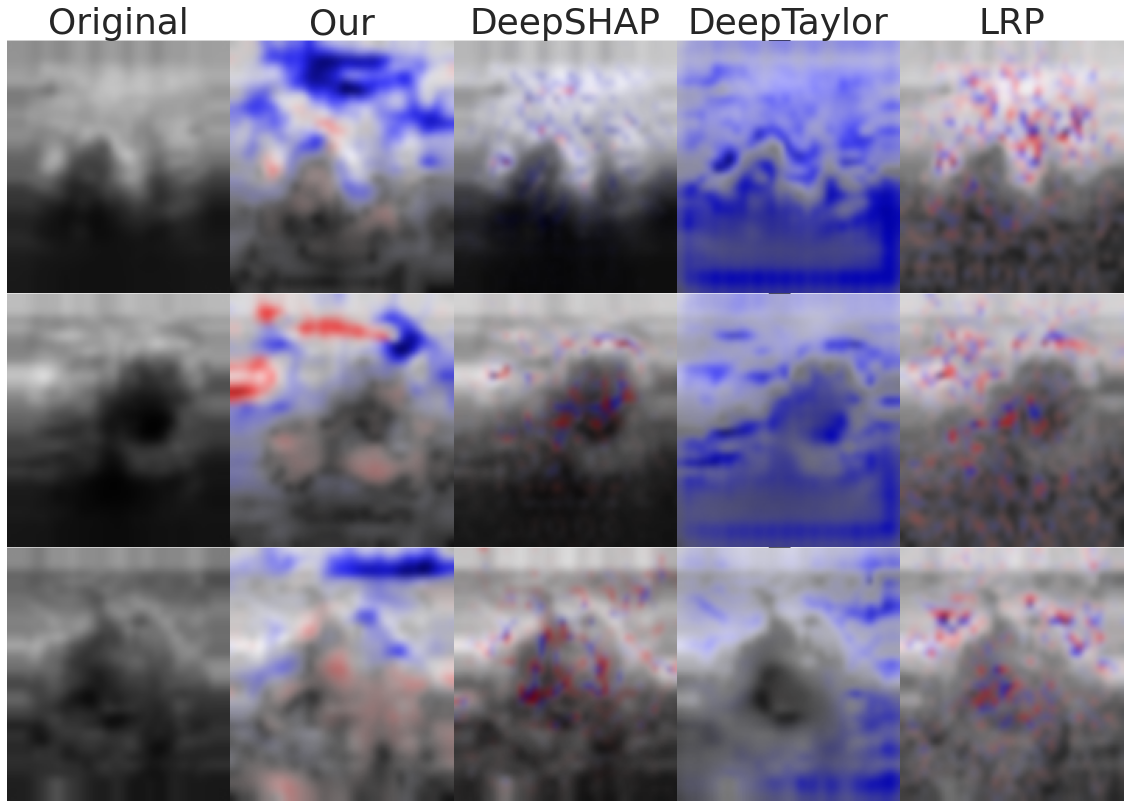}\hspace{1em}
	\includegraphics[width=.45\textwidth]{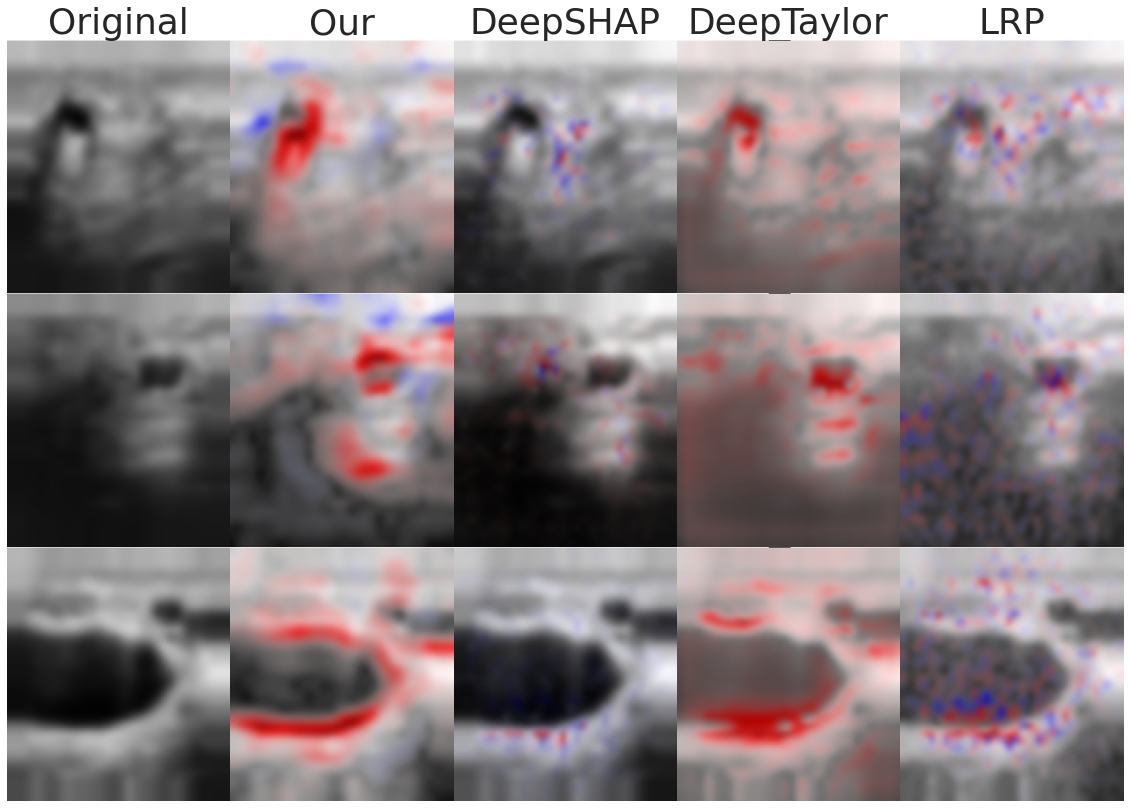}
	\caption{\modified{}Example images for benign (left) and malignant (right) breast lesions with blue/red indicating benignity/malignancy-indicating regions on the BreastMNIST dataset.
	}
	\label{Fig:Visualization_BreastMNIST}
\end{figure}

\begin{figure}
	\centering
	\hspace{-1em}\includegraphics[width=.48\textwidth]{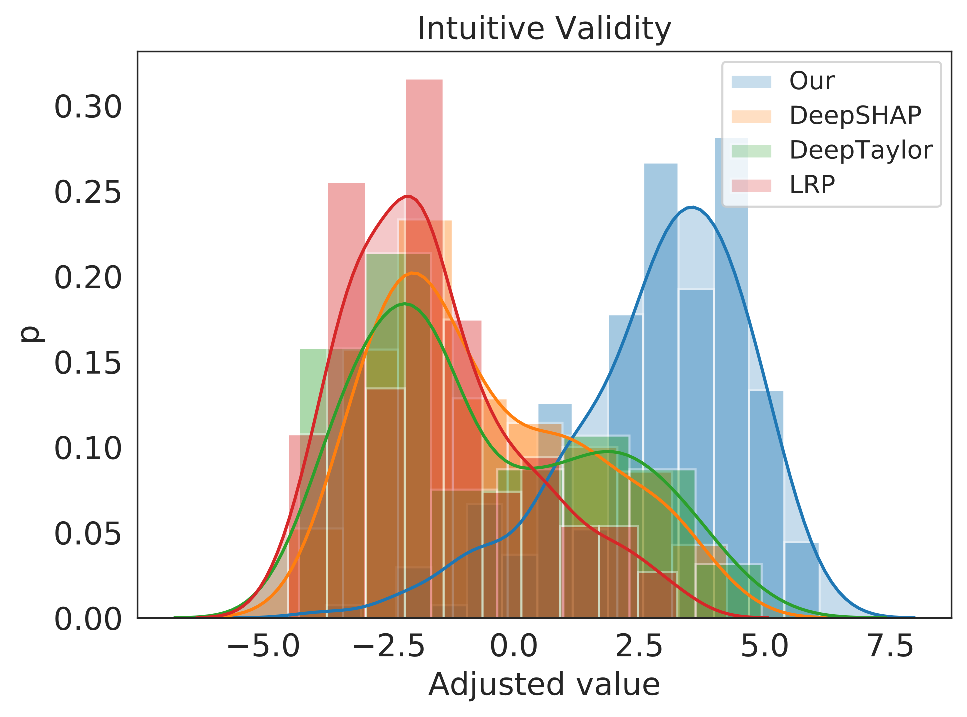}
	\raisebox{.4em}{\includegraphics[width=.48\textwidth]{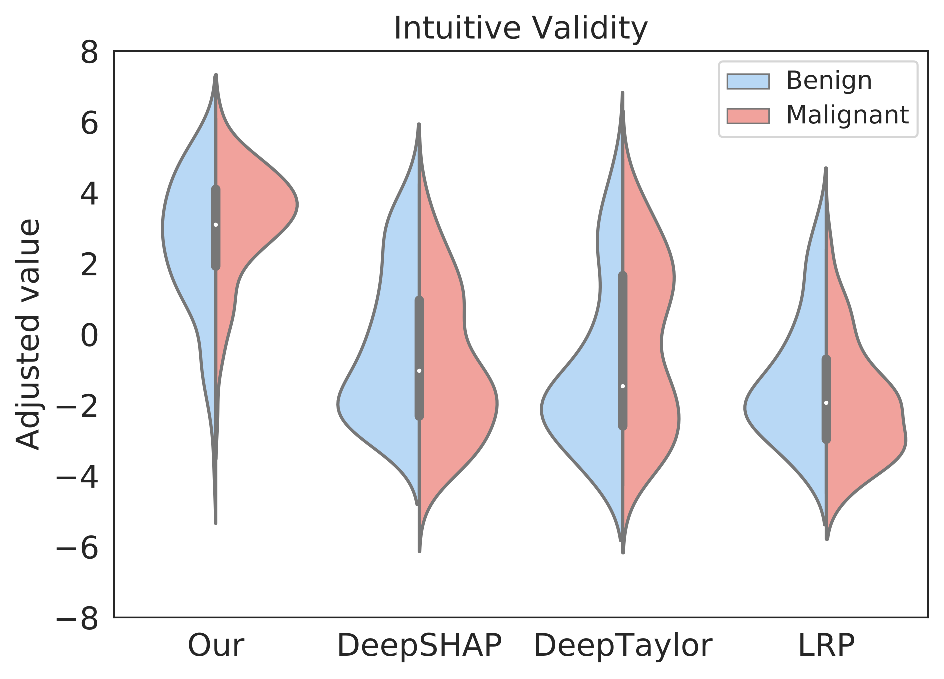}}\\
	\hspace{-1em}\includegraphics[width=.48\textwidth]{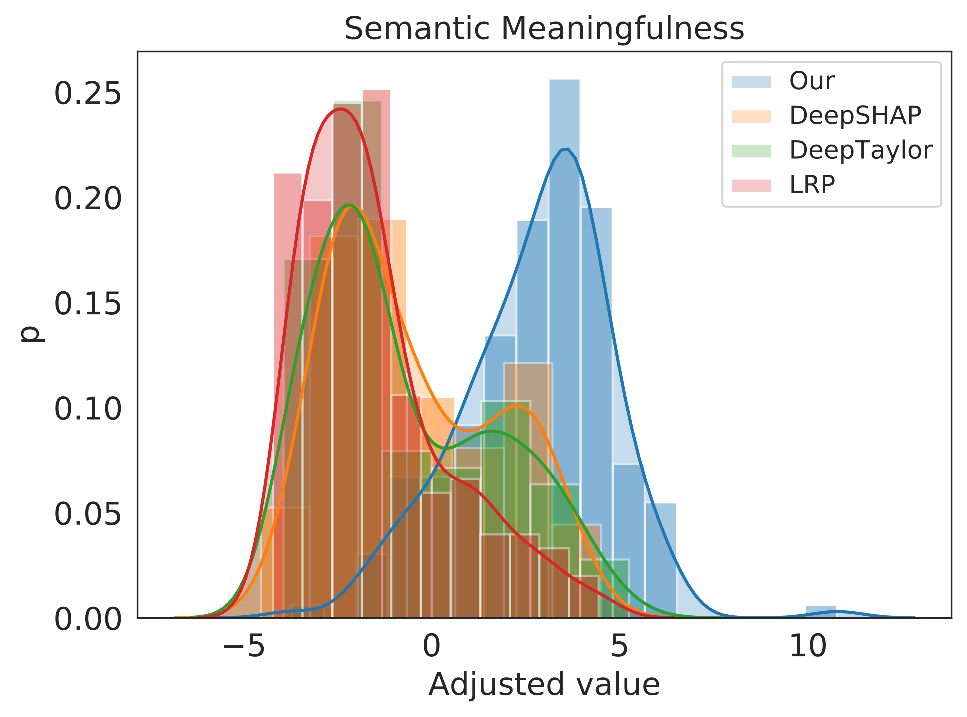}
	\raisebox{.4em}{\includegraphics[width=.48\textwidth]{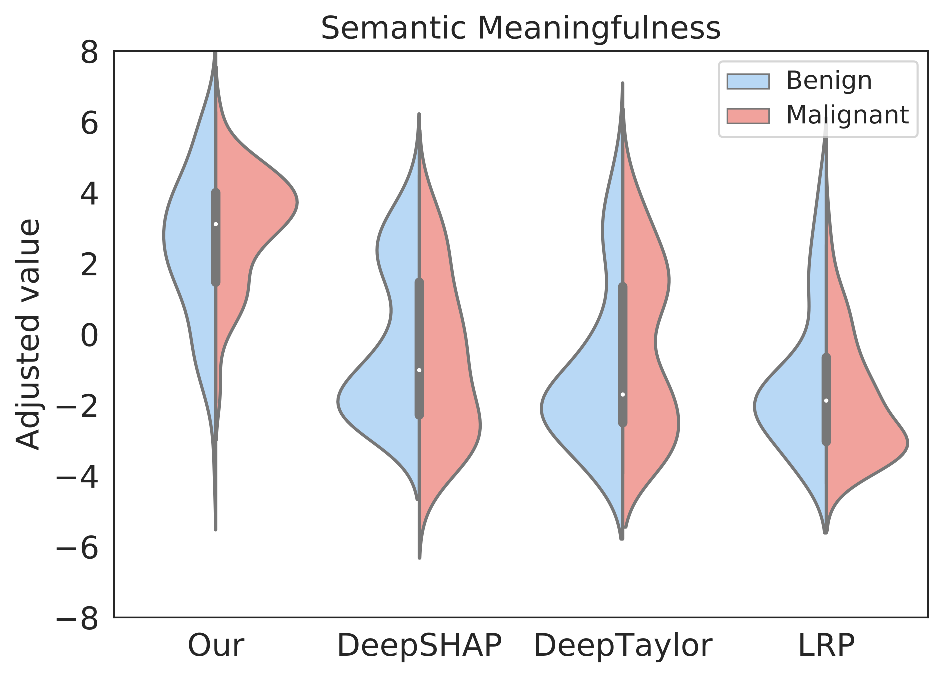}}\\
	\hspace{-1em}\includegraphics[width=.48\textwidth]{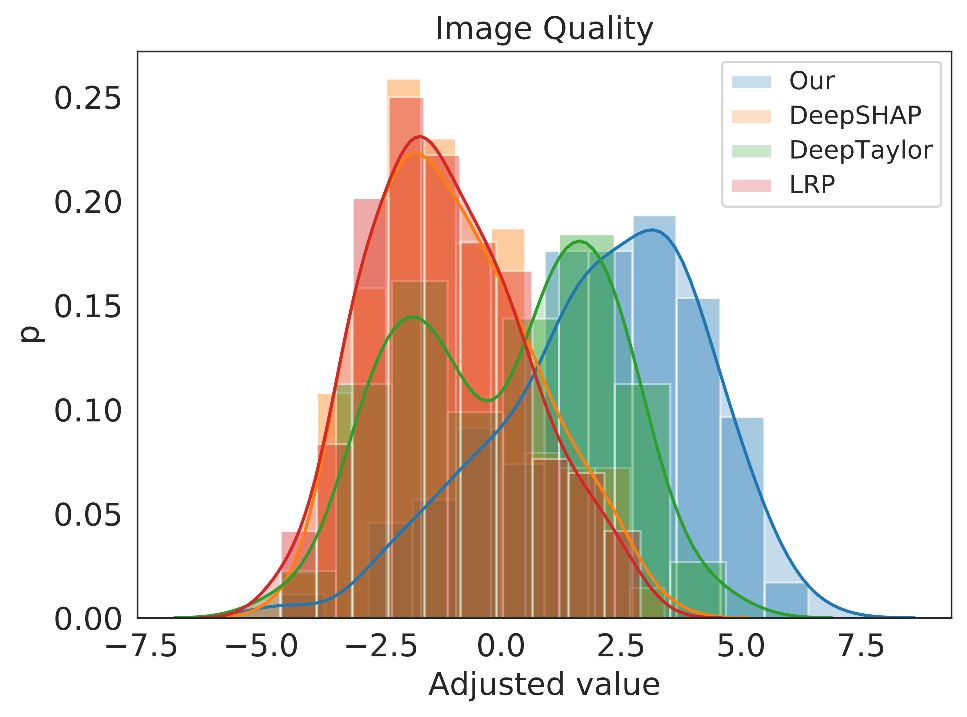}
	\raisebox{.4em}{\includegraphics[width=.48\textwidth]{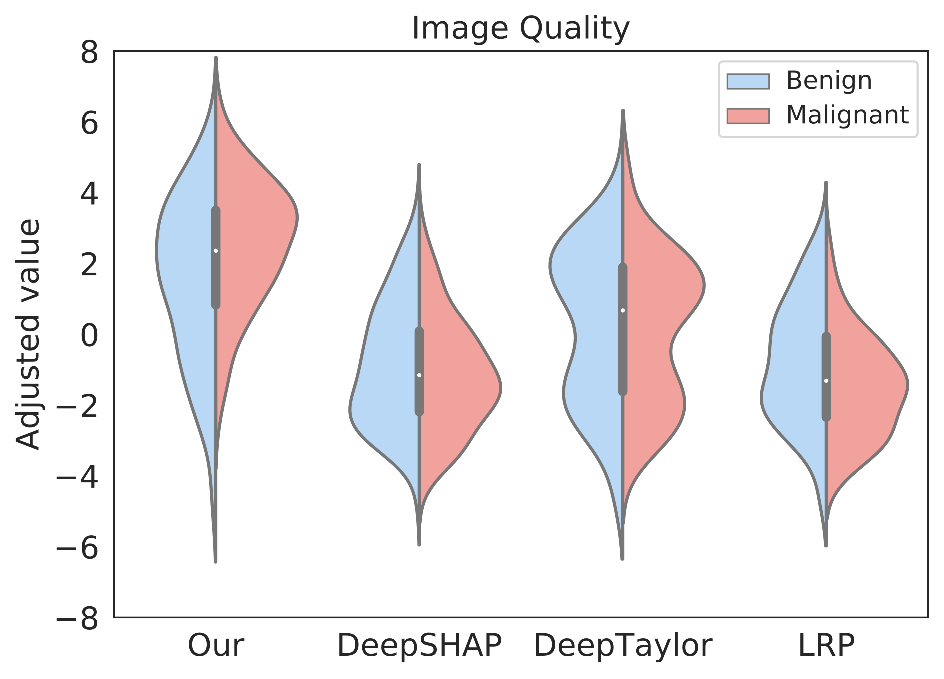}}

	\caption{{\modified{}Examplary comparison of the z-adjusted questionnaire results for each criterion and method on the LIDC-IDRI dataset, cumulative (left) and by lesion type (right), for benign (blue) and malignant (red) lesions.}}
	\label{Fig:US} %
\end{figure}

\begin{table}[]
\centering
\begin{tabular}{cccc}
\toprule
           & \textbf{Intuitivity}               & \textbf{Semantics}                 & \textbf{Quality}                  \\
\midrule
\midrule
\multicolumn{4}{c}{Questionnaire results (raw)}\\
\midrule
\textbf{Ours}       & \textbf{1.92}~~ {[}~0.50,~3.13{]}      & \textbf{1.76}~~ {[}~0.25,~3.13{]}      & \textbf{1.04}~~ {[}-0.63,~2.50{]}    \\
DeepSHAP   & -1.56*~ {[}-3.13,~0.13{]}   & -1.54*~ {[}-3.13,~0.25{]}   & -2.02*~ {[}-3.25,-0.63{]} \\
DeepTaylor & -1.52*~ {[}-3.25,~0.38{]}   & -1.72*~ {[}-3.38,~0.13{]}   & -0.85~~ {[}-2.50,~0.75{]} \\
LRP        & -2.53** {[}-3.63,-1.25{]} & -2.57** {[}-3.75,-1.13{]} & -2.16*~ {[}-3.38,-0.75{]}\\
\midrule
\multicolumn{4}{c}{Questionnaire results (z-adjusted)}\\
\midrule%
\textbf{Ours}       & \textbf{2.84}~~ {[}~1.50,~3.98{]}      & \textbf{2.78}~~ {[}~1.39,~4.10{]}      & \textbf{2.04}~~ {[}~0.50,~3.41{]}      \\
DeepSHAP   & -0.64*~ {[}-2.00,~0.86{]}   & -0.52*~ {[}-1.97,~1.05{]}   & -1.02*~ {[}-2.12,~0.16{]}   \\
DeepTaylor & -0.60*~ {[}-2.16,~1.10{]}   & -0.71*~ {[}-2.23,~1.00{]}   & 0.14~~ {[}-1.32,~1.57{]}     \\
LRP        & -1.61** {[}-2.72,-0.33{]} & -1.55** {[}-2.73,-0.11{]} & -1.16*~ {[}-2.24,-0.02{]}\\
\midrule
\multicolumn{4}{c}{Average rank}\\
\midrule%
\textbf{Ours}       & \textbf{1.28}~~~ {[}1.00,1.75{]}    & \textbf{1.36}~~ {[}1.00,1.88{]}   & \textbf{1.52}~ {[}1.06,2.19{]}  \\
DeepSHAP   & 2.78**~ {[}2.19,3.38{]}  & 2.75*~ {[}2.13,3.31{]}  & 3.04* {[}2.50,3.50{]} \\
DeepTaylor & 2.71*~~ {[}2.06,3.31{]}   & 2.75*~ {[}2.06,3.38{]}  & 2.36~ {[}1.75,3.00{]}  \\
LRP        & 3.23*** {[}2.75,3.63{]} & 3.14** {[}2.63,3.56{]} & 3.07* {[}2.50,3.56{]}\\
\bottomrule
\end{tabular}
\caption{\label{Tab:UserStudy_LIDC}{\modified Average questionnaire results (higher is better) and rank (lower is better)} per algorithm {\modified{}on the LIDC-IDRI dataset} with 95\,\% CI. $p$-values for two-tailed $t$-test with $t(6)$ are indicated as $p<.05$*, $p<.01$**, $p<.001$***. The best result is marked bold.}
\end{table}

\begin{table}[]
\centering
\begin{tabular}{cccc}
\toprule
           & \textbf{Intuitivity}               & \textbf{Semantics}                 & \textbf{Quality}                  \\
\midrule
\midrule
\multicolumn{4}{c}{Questionnaire results (raw)}\\
\midrule
\textbf{Ours}       & \textbf{~1.45}~~ {[}~0.36,~2.45{]}      & \textbf{~1.33}~~ {[}~0.27,~2.36{]}      & \textbf{~1.72}~~ {[}~0.64,~2.64{]}    \\
DeepSHAP   & -1.77** {[}-2.82,-0.64{]}   & -1.95** {[}-3.00,-0.82{]}   & -1.42** {[}-2.55,-0.27{]} \\
DeepTaylor & ~0.91~~ {[}-0.55,~2.18{]}   & ~0.90~~ {[}-0.55,~2.18{]}   & ~0.75~~ {[}-0.64,~2.00{]}   \\
LRP        & -1.41** {[}-2.55,-0.18{]} & -1.63** {[}-2.73,-0.45{]} & -1.19** {[}-2.36,~0.00{]}\\
\midrule
\multicolumn{4}{c}{Questionnaire results (z-adjusted)}\\
\midrule
\textbf{Ours}       & \textbf{~1.66}~~ {[}~0.65,~2.57{]}      & \textbf{1.67}~~ {[}~0.63,~2.62{]}      & \textbf{1.76}~~ {[}~0.70,~2.71{]} \\
DeepSHAP   & -1.56** {[}-2.63,~0.39{]}   & -1.62** {[}-2.70,-0.43{]}   & -1.38** {[}-2.47,-0.20{]}   \\
DeepTaylor & ~1.11~~ {[}-0.26,~2.37{]}   & ~1.24~~ {[}-0.12,~2.49{]}   & 0.78~~ {[}-0.53,~1.99{]}     \\
LRP        & -1.21** {[}-2.33,~0.02{]} & -1.29** {[}-2.41,-0.07{]} & -1.15** {[}-2.29,~0.04{]}\\
\midrule
\multicolumn{4}{c}{Average rank}\\
\midrule
\textbf{Ours}       & \textbf{1.75}~~ {[}1.36,2.23{]}    & \textbf{1.77}~~ {[}1.32,2.32{]}   & \textbf{1.64}~~ {[}1.23,2.14{]}  \\
DeepSHAP   & 3.26** {[}2.82,3.64{]}  & 3.23** {[}2.77,3.64{]}  & 3.12** {[}2.59,3.59{]} \\
DeepTaylor & 1.99~~ {[}1.41,2.64{]}   & 1.99~~ {[}1.45,2.64{]}  & 2.27~~ {[}1.73,2.86{]}  \\
LRP        & 3.00** {[}2.45,3.45{]} & 3.02*~ {[}2.50,3.45{]} & 2.97*~ {[}2.36,3.50{]}\\
\bottomrule
\end{tabular}
\caption{\label{Tab:UserStudy_Breast}\modified{}Average questionnaire results (higher is better) and rank (lower is better) per algorithm on the BreastMNIST dataset with 95\,\% CI. $p$-values for two-tailed $t$-test with $t(6)$ are indicated as $p<.05$*, $p<.01$**. The best result is marked bold.}
\end{table}

\begin{table}[]
\centering
\begin{tabular}{cccc}
\toprule
           & \textbf{Intuitivity}               & \textbf{Semantics}                 & \textbf{Quality}                  \\
\midrule
\midrule
\multicolumn{4}{c}{Questionnaire results (raw)}\\
\midrule
Ours       & ~0.67~ {[}-0.67,~1.89{]}      & ~0.72~ {[}-0.78,~2.11{]}      & -0.12 {[}-1.44,~1.22{]}    \\
DeepSHAP   & -1.41~ {[}-2.78,~0.00{]}   & -1.48~ {[}-2.89,~0.00{]}   & -0.23 {[}-2.00,~1.44{]} \\
\textbf{DeepTaylor} & \textbf{~1.91}~ {[}~0.56,~3.00{]}   & \textbf{~2.16}~ {[}~0.78,~3.22{]}   & \textbf{~0.70} {[}-0.78,~2.11{]}   \\
LRP        & -1.81* {[}-3.00,-0.44{]} & -1.91* {[}-3.22,-0.44{]} & -1.45 {[}-2.56,~0.33{]}\\
\midrule
\multicolumn{4}{c}{Questionnaire results (z-adjusted)}\\
\midrule
Ours       & ~0.83~ {[}-0.47,~2.03{]}      & ~0.85~ {[}-0.58,~2.13{]}      & ~0.08 {[}-1.18,~1.36{]}      \\
DeepSHAP   & -1.25~ {[}-2.60,~0.16{]}   & -1.35~ {[}-2.71,~0.07{]}   & -0.03 {[}-1.83,~1.69{]}   \\
\textbf{DeepTaylor} & \textbf{~2.07}~ {[}~0.75,~3.12{]}   & \textbf{~2.29}~ {[}~0.96,~3.32{]}   & \textbf{~0.90} {[}-0.58,~2.28{]}     \\
LRP        & -1.65* {[}-2.88,-0.32{]} & -1.79* {[}-3.07,-0.36{]} & -0.95 {[}-2.39,~0.48{]}\\
\midrule
\multicolumn{4}{c}{Average rank}\\
\midrule
Ours       & 2.13~ {[}1.61,2.72{]}    & 2.15~ {[}1.61,2.72{]}   & 2.61 {[}1.94,3.27{]}  \\
DeepSHAP   & 3.04~ {[}2.50,3.50{]}  & 3.06~ {[}2.50,3.56{]}  & 2.30 {[}1.61,3.00{]} \\
\textbf{DeepTaylor} & \textbf{1.52}~ {[}1.06,2.17{]}   & \textbf{1.50}~ {[}1.06,2.11{]}  & \textbf{2.16} {[}1.50,2.89{]}  \\
LRP        & 3.30* {[}2.83,3.72{]} & 3.29* {[}2.78,3.67{]} & 2.92 {[}2.39,3.44{]}\\
\bottomrule
\end{tabular}
\caption{\label{Tab:UserStudy_CifarCD}\modified{}Average questionnaire results (higher is better) and rank (lower is better) per algorithm on the CIFAR-10 cats-vs.-dogs dataset with 95\,\% CI. $p$-values for two-tailed $t$-test with $t(7)$ are indicated as $p<.05$*. The best result is marked bold.}
\end{table}

\begin{table}[]
\centering
\begin{tabular}{cccc}
\toprule
           & \textbf{Intuitivity}               & \textbf{Semantics}                 & \textbf{Quality}                  \\
\midrule
\midrule
\multicolumn{4}{c}{Questionnaire results (raw)}\\
\midrule
Ours       & ~0.95~ {[}-0.67,~2.33{]}      & ~1.49~~ {[}~0.00,~2.67{]}      & ~0.49 {[}-1.00,~1.83{]}    \\
DeepSHAP   & -1.78* {[}-3.33,~0.00{]}   & -1.97*~ {[}-3.50,-0.17{]}   & -1.51 {[}-3.33,~0.33{]} \\
\textbf{DeepTaylor} & \textbf{~2.43}~ {[}~0.66,~3.50{]}   & \textbf{~2.70}~~ {[}~1.17,~3.67{]}   & \textbf{~1.75} {[}-0.17,~3.33{]}   \\
LRP        & -2.20* {[}-3.66,-0.33{]} & -2.46** {[}-3.83,-0.83{]} & -1.87 {[}-3.50,-0.00{]}\\
\midrule
\multicolumn{4}{c}{Questionnaire results (z-adjusted)}\\
\midrule
Ours    & ~1.10~ {[}-0.28,~2.30{]}      & ~1.55~~ {[}~0.33,~2.54{]}      & ~0.77~ {[}-0.67,~2.10{]}      \\
DeepSHAP   & -1.64* {[}-3.30,~0.18{]}   & -1.91*~ {[}-3.54,-0.04{]}   & -1.22* {[}-3.01,~0.67{]}   \\
\textbf{DeepTaylor} & \textbf{~2.58}~ {[}~1.25,~3.58{]}   & \textbf{~2.76}~~ {[}~1.53,~3.60{]}   & \textbf{~2.03}~ {[}~0.31,~3.51{]}     \\
LRP        & -2.05* {[}-3.65,-0.18{]} & -2.40** {[}-3.78,-0.72{]} & -1.58* {[}-3.27,~0.30{]}\\
\midrule
\multicolumn{4}{c}{Average rank}\\
\midrule
Ours       & 2.12~ {[}1.67,2.67{]}    & 1.94~~ {[}1.58,2.33{]}   & 2.39 {[}1.75,3.08{]}  \\
DeepSHAP   & 3.16~ {[}2.67,3.58{]}  & 3.24*~ {[}2.75,3.58{]}  & 2.83 {[}2.00,3.50{]} \\
\textbf{DeepTaylor} & \textbf{1.32}~ {[}1.00,2.00{]}   & \textbf{1.31}~~ {[}1.00,1.92{]}  & \textbf{1.72} {[}1.08,2.50{]}  \\
LRP        & 3.40* {[}2.83,3.83{]} & 3.51** {[}3.17,3.83{]} & 3.05 {[}2.33,3.67{]}\\
\bottomrule
\end{tabular}
\caption{\label{Tab:UserStudy_CifarTC}\modified{}Average questionnaire results per algorithm (higher is better) and rank (lower is better) on the CIFAR-10 trucks-vs.-cars dataset with 95\,\% CI. $p$-values for two-tailed $t$-test with $t(4)$ are indicated as $p<.05$*, $p<.01$**. The best result is marked bold.}
\end{table}

The results of the user study are {\modified listed} in {\modified{}Tables~\ref{Tab:UserStudy_LIDC} through \ref{Tab:UserStudy_CifarTC}} and visualized {\modified exemplarily for the LIDC-IDRI data set} in Fig.~\ref{Fig:US}. A qualitative comparison of our approach to DeepSHAP, DeepTaylor and LRP is depicted in Fig.~\ref{Fig:Visualization}. %
{\modified{}
\paragraph{LIDC-IDRI}~
}
Regarding the tested criteria, our method outperformed all other tested methods {\modified{}on the LIDC-IDRI dataset} with average raw values of 1.92, 95\,\% CI [0.50, 3.13], 1.76 [0.25, 3.13], and 1.04 [-0.63, 2.50] for intuitive validity, semantic meaningfulness and image quality, respectively. Average values for the second best method for each criterion were -1.52 (DeepTaylor, $t(6)=3.07, p=.018$, two-tailed $t$-test), -1.54 (DeepSHAP, $t(6)=3.26, p=.014$), and -.85 (DeepTaylor, $t(6)=1.71, n.s.$). Regarding intuitive validity and semantic meaningfulness, our method was significantly superior to all other tested methods with $p<.05$. Regarding image quality, significant superiority could be shown to DeepSHAP and LRP ($p=.015/.015)$, while for DeepTaylor the result was non-significant with $p=.132$.

The z-adjusted values of our method were 2.84, 95\,\% CI [1.50,3.98], 2.78 [1.39,4.10] and 2.04 [0.50,3.41] for intuitive validity, semantic meaningfulness and image quality. Significant superiority could be shown analogously to the unadjusted values to each method and each criterion with $p$-values between $.002-.019$ with the exception of image quality from the DeepTaylor algorithm ($t(6)=1.71, p=.131$).

The average rank of our method in direct comparison was 1.28, 1.36 and 1.52, for intuitive validity, semantic meaningfulness and image quality, compared with 2.78, 2.71, 3.23 (DeepSHAP/DeepTaylor/LRP) for intuitive validity, 2.75, 2.75, 3.41 for semantic meaningfulness and 3.04, 2.36, 3.07 for image quality. $p-values$ ranged from $<.001-.019$, again with the exception of image quality for the DeepTaylor algorithm ($p=.152$).

\begin{figure}
\centering
\includegraphics[width=.49\textwidth]{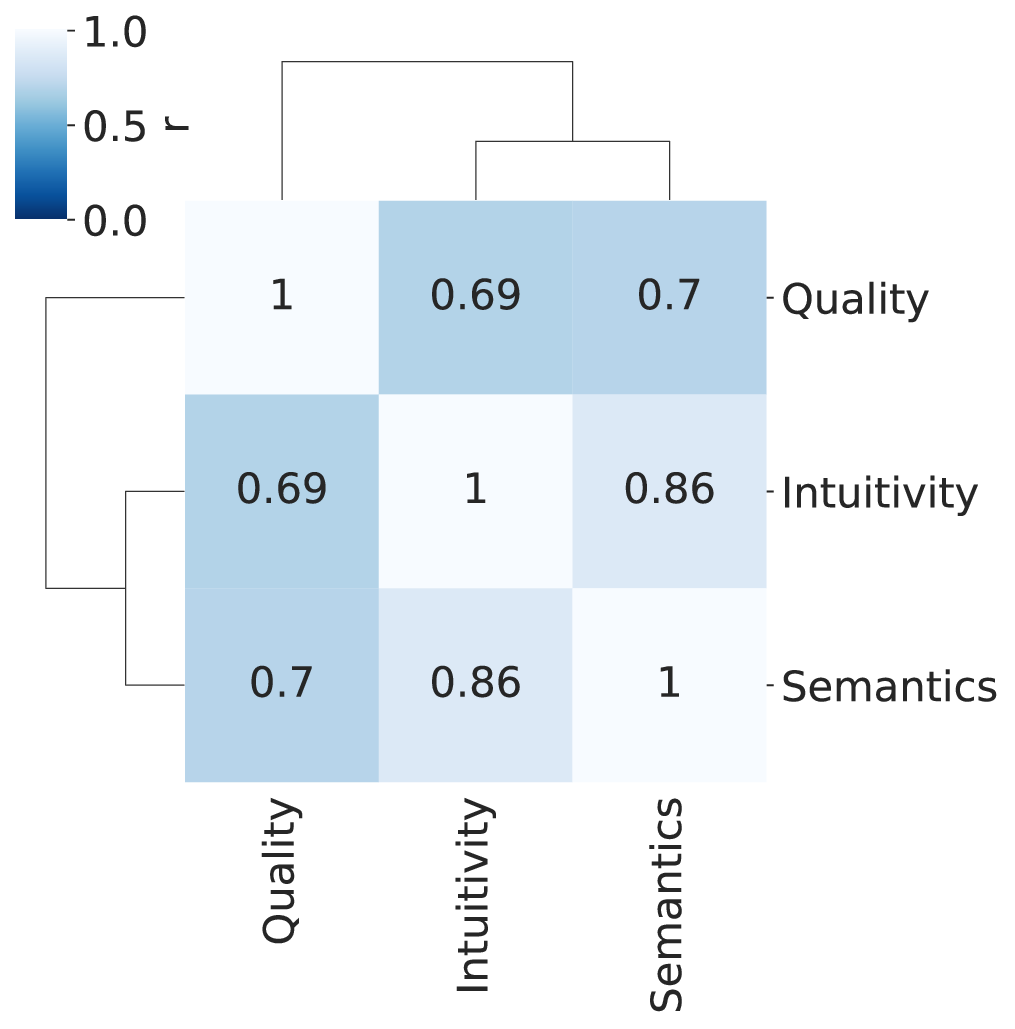}\includegraphics[width=.49\textwidth]{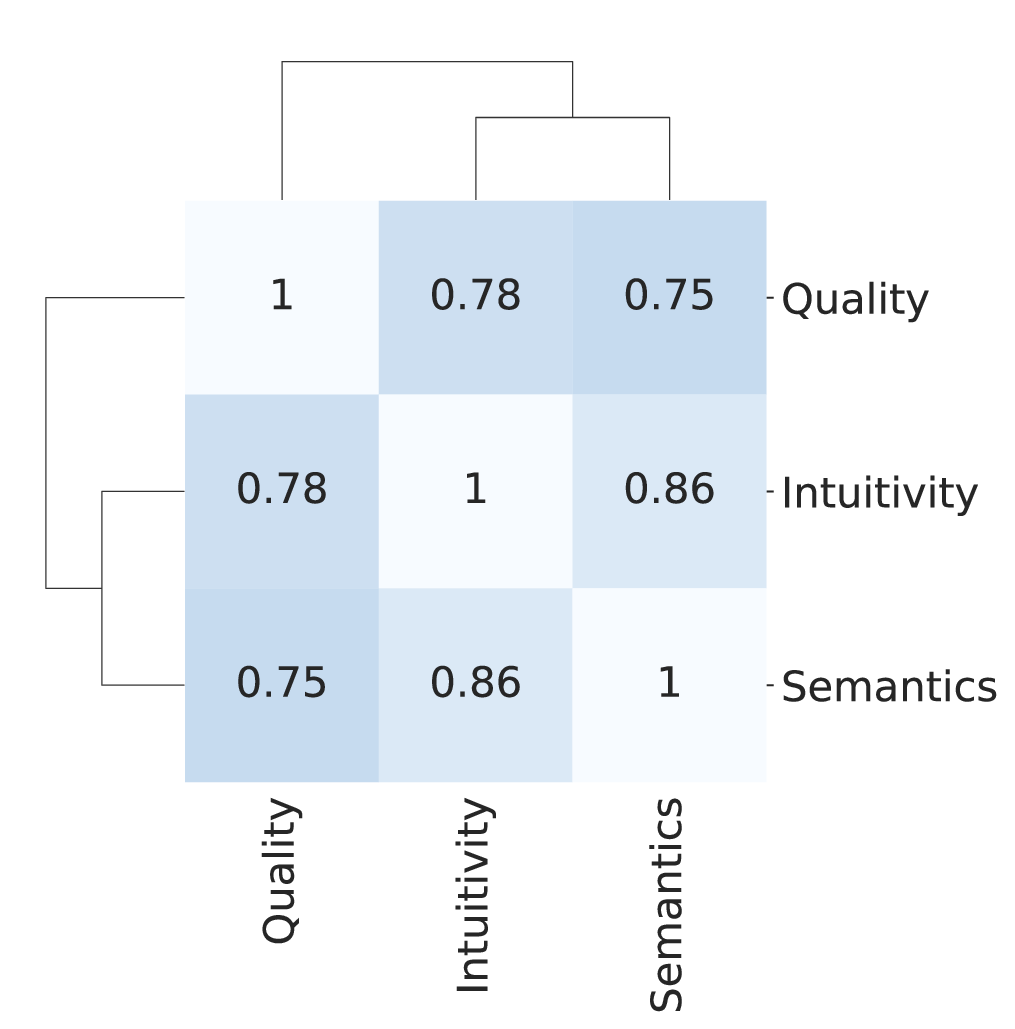}\\
\includegraphics[width=.49\textwidth]{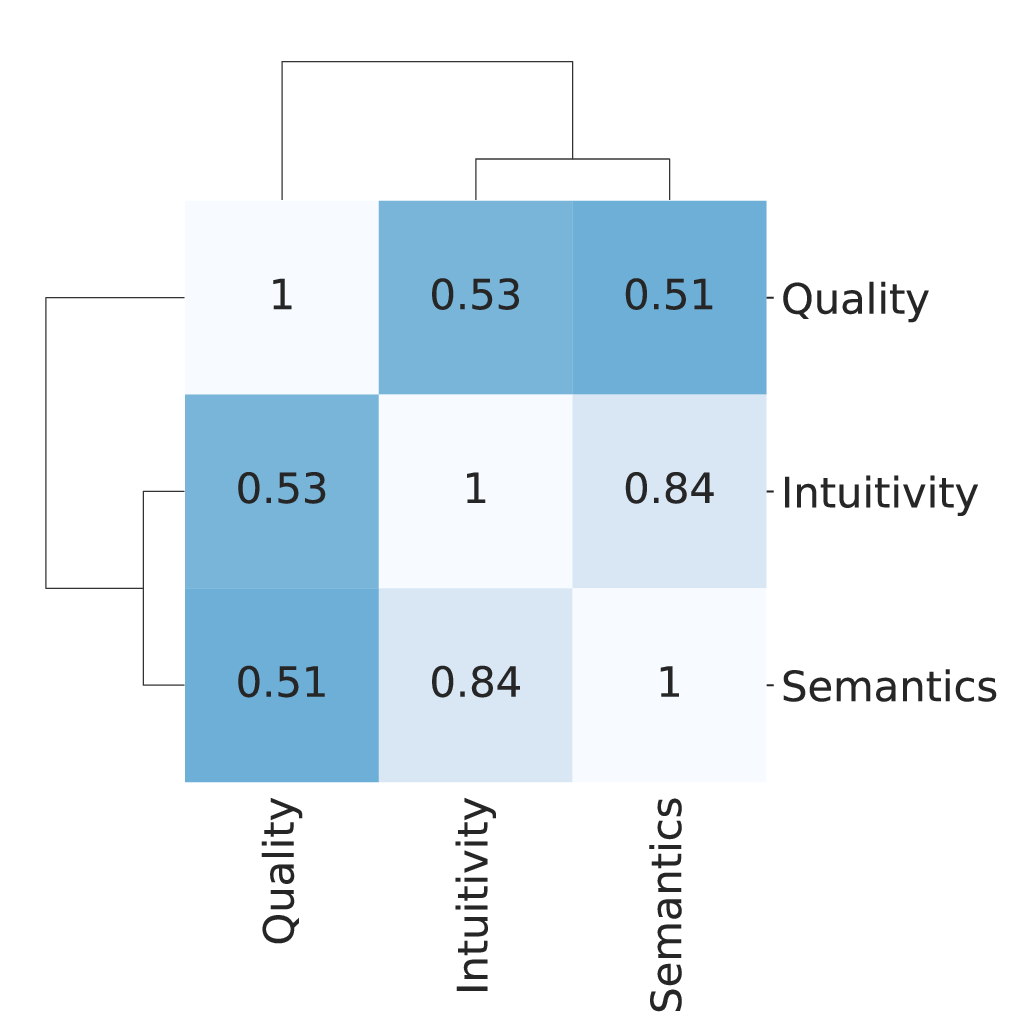}\includegraphics[width=.49\textwidth]{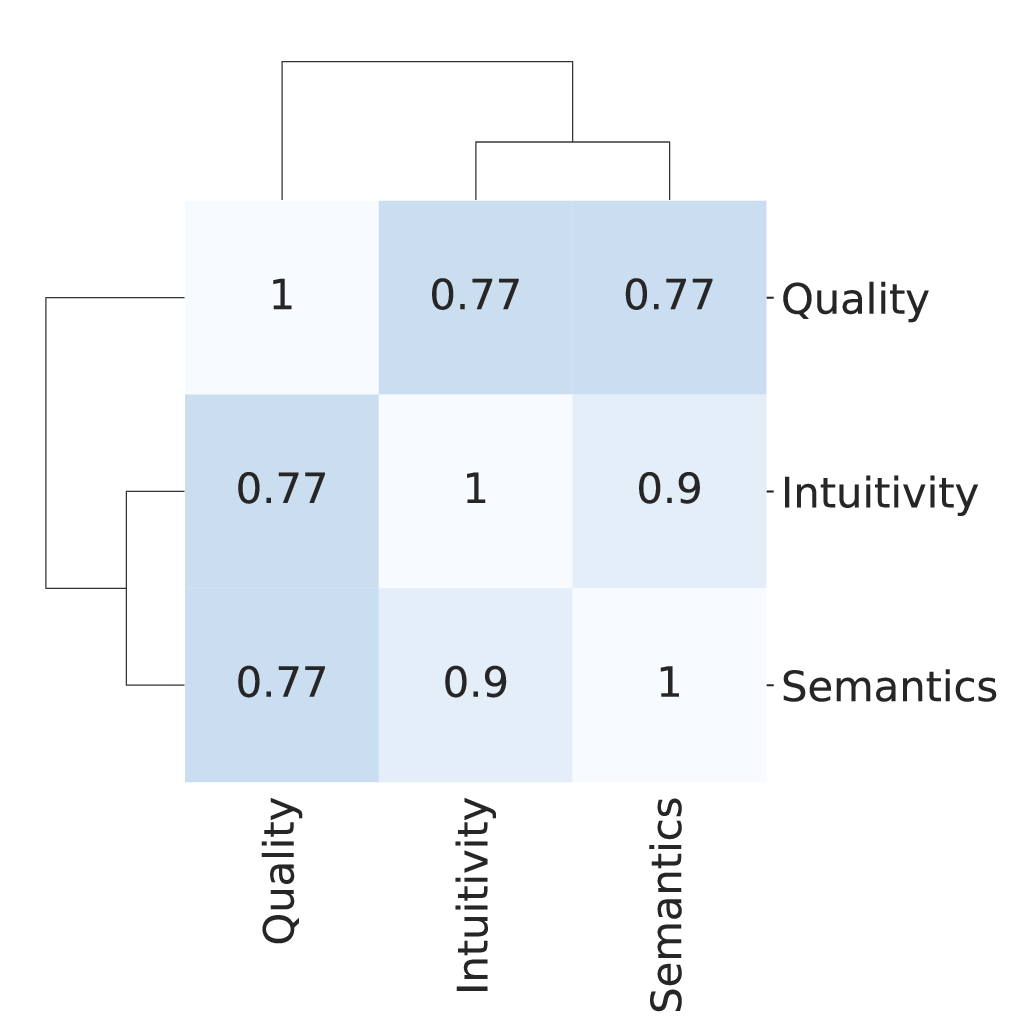}
\caption{Dendrogram of the pairwise Pearson correlation {\modified{}matrices} for the analyzed criteria {\modified{}on the LIDC-IDRI (top-left), BreastMNIST (top-right), cats-vs.-dogs (bottom-left) and trucks-vs.-cars (bottom-right) datasets.}}
\label{Fig:CriteriaCorr}
\end{figure}

\begin{figure}[h!]
\centering

    \raisebox{-.2em}{\includegraphics[width=.49\textwidth]{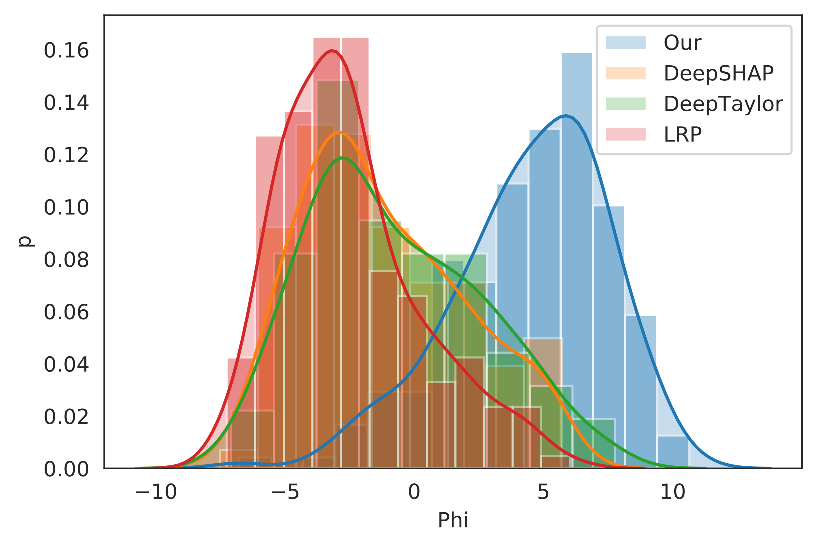}}
    \includegraphics[width=.49\textwidth]{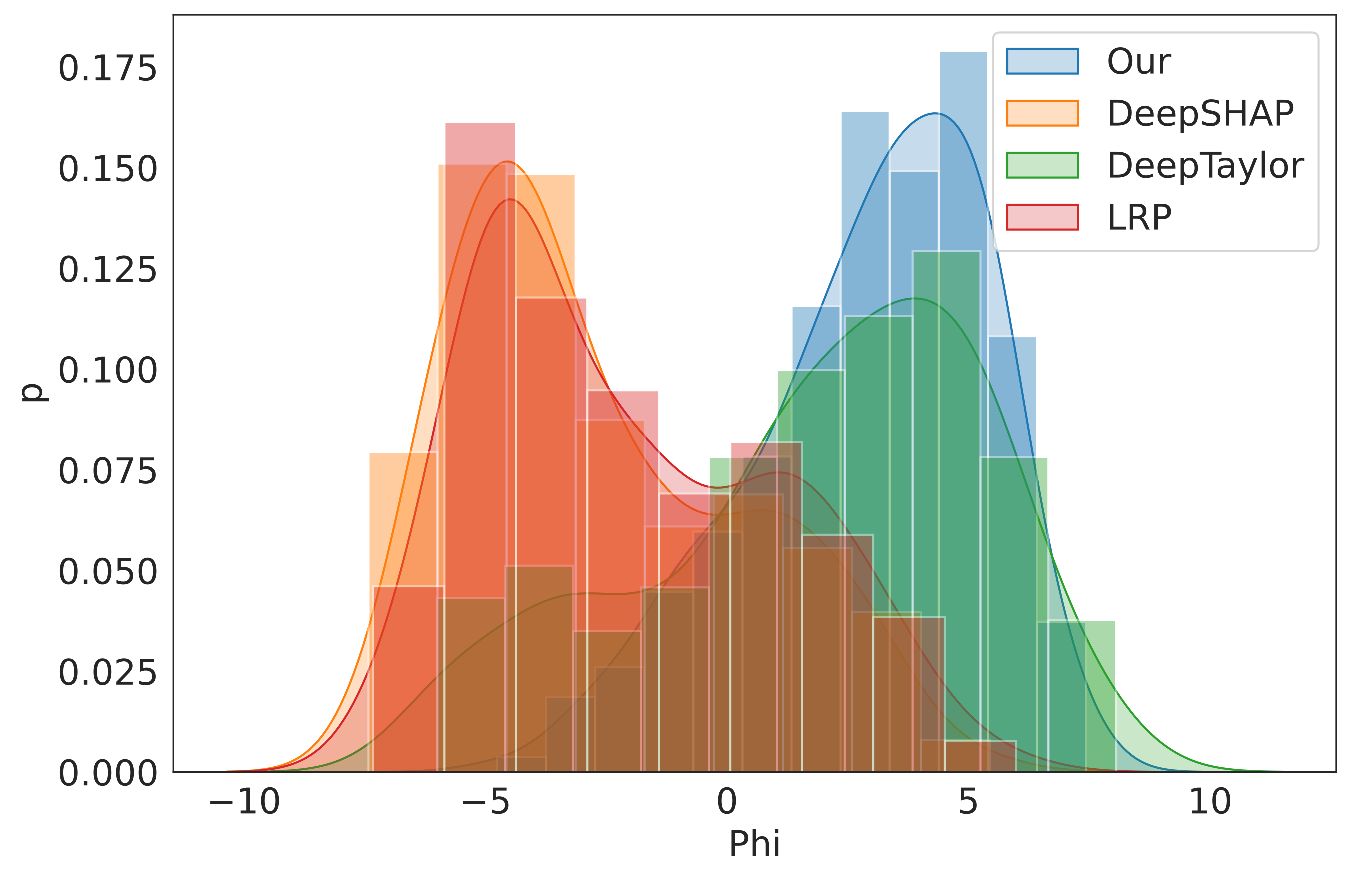}\\
    \includegraphics[width=.49\textwidth]{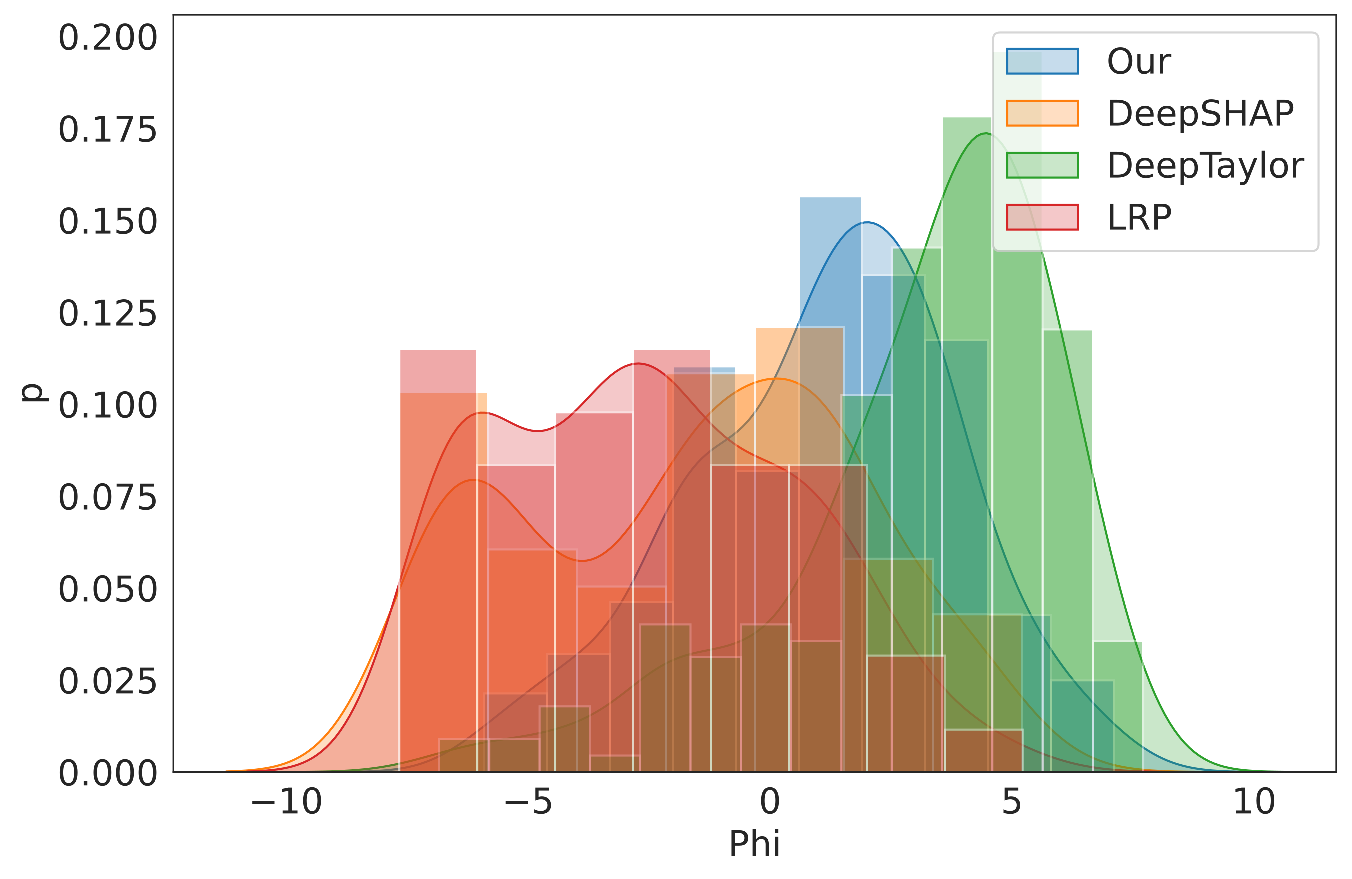}
    \includegraphics[width=.49\textwidth]{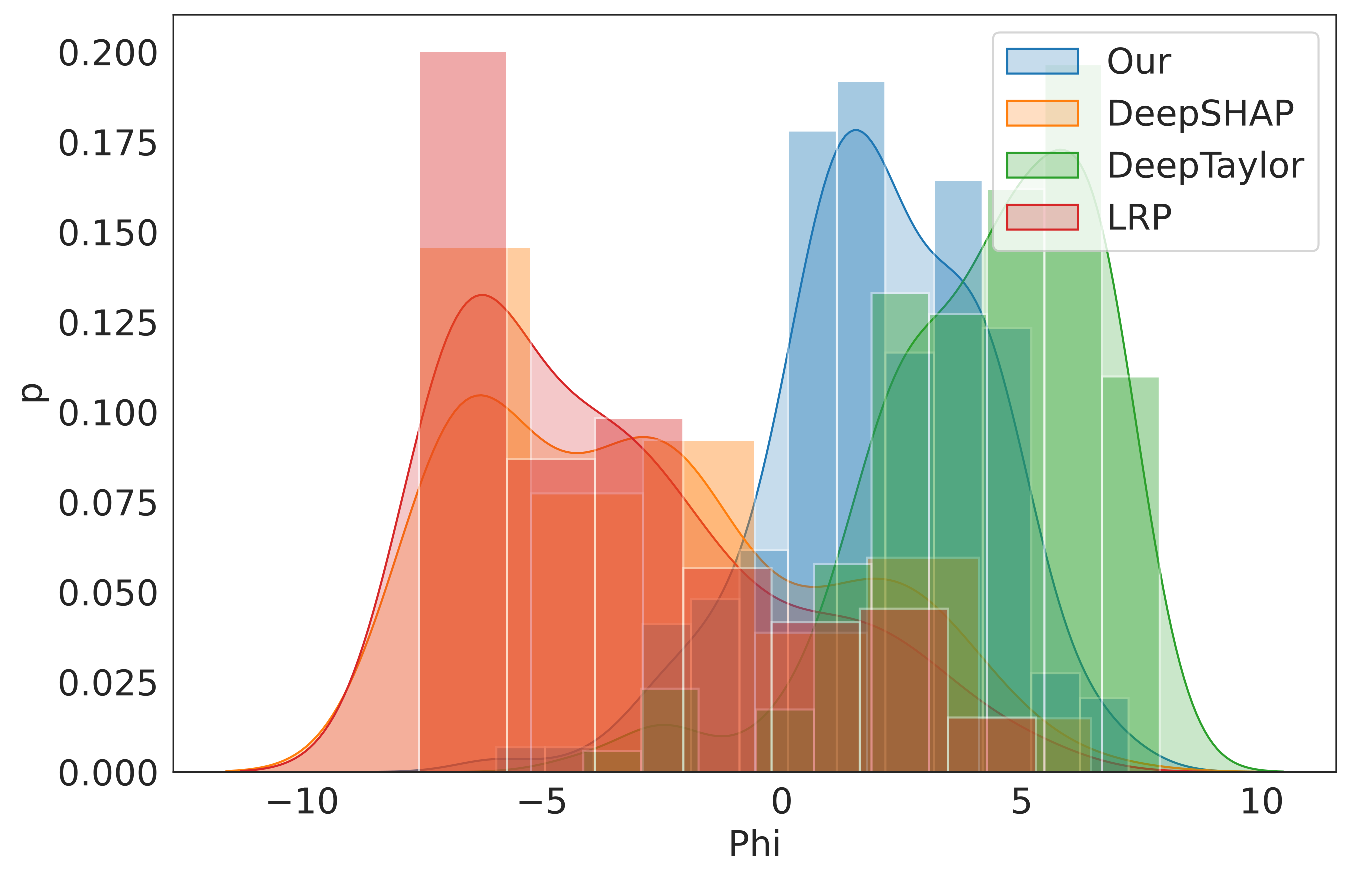}\\
    \vspace{1em}
    \begin{tabular}{ccccc}\toprule
             & LIDC-IDRI & BreastMNIST & Cats-vs.-Dogs & Trucks-vs.-Cars\\
             \midrule
             \midrule
             \multicolumn{5}{c}{$\Phi$ coefficients}\\
             \midrule
             Intuitivity & .358 & .341 & .356 & .340\\
             Semantics & .367 & .343 & .377 & .357\\
             Quality & .274 & .316 & .267 & .303\\
             \bottomrule
    \end{tabular}

\caption{\modified{}Comparison of the analyzed approaches when extracting a general factor $\Phi$ using PCA and its L1-normalized largest eigenvector. $\Phi$ accounts for 84.6/86.5/76.6/88.0\,\% of the observed variance on the LIDC-IDRI (top-left), the BreastMNIST (top-right), the Cats-vs.-Dogs (bottom-left) and the Trucks-vs.-Cars (bottom-right) dataset. The table shows the composition of $\Phi$ on each dataset.}
\label{Fig:GeneralFactor}
\end{figure}

The pairwise Pearson correlations between the tested criteria over all questionnaires were between $r=.689$ (image quality and intuitive validity) and $r=.862$ (intuitive validity and semantic meaningfulness), implying an expected association between the tested criteria (see Fig.~\ref{Fig:CriteriaCorr}). The inter-observer reliability was $\rho=.639/.590/.557$ for intuitive validity, semantic meaningfulness and image quality, respectively, implying a substantial to moderate inter-observer agreement \cite{landis1977measurement}. The Kaiser-Mayer-Olkin criterion for our sample was $KMO=.716$, implying the adequacy of our distribution for principal component analysis \cite{dziuban1974correlation}. Applying the PCA, it was possible to extract a general factor of preferability $\Phi$, which accounted for 84.6\,\% of the observed variance (see Fig.~\ref{Fig:GeneralFactor}). 

The differences between the questionnaire sheets A and B for each method and criterion were all non-significant with $p$-values between .133 and .957 (avg: .649, $.002<t(6)<1.68$), implying that there was no systematic difference between the questionnaires and {\modified{}that} no order effects occurred.

{\modified{}
\paragraph{BreastMNIST}~
On the BreastMNIST datasets our method reached average raw values of 1.45, 1.33 and 1.72 for intuitive validity, semantic meaningfulness and image quality, respectively, reaching the highest values amongst the tested algorithms, second highest was DeepTaylor with 0.91/0.90/0.75. While the results were significantly superior to DeepSHAP and LRP ($t(9)=3.34-4.07$, all $p<.01$), the difference to DeepTaylor was not significant ($0.56<t(9)<1.73, p>.05$), with the same pattern showing for the z-adjusted values. The average rank amongst the methods was 1.75, 1.77 and 1.64 for our approach with DeepTaylor being second with 1.99, 1.99 and 2.27, respectively. Again, results were significantly superior to DeepSHAP and LRP ($2.96<t(9)<3.97$, all $p<.05$) and not significant for DeepTaylor ($0.51<t(9)<1.84$). The inter-observer reliability was between .386 and .455, implying a fair to moderate correlation. The Kaiser-Mayer-Olkin criterion was $\mathrm{KMO}=.742$ and the largest principal component was responsible for 86.5\,\% of the variance.

\paragraph{CIFAR-10}~
On the cats-vs.-dogs dataset, DeepTaylor was clearly superior to the other tested methods with average z-adjusted values of 2.07, 2.29 and 0.90 for intuitive validity, semantic meaningfulness and image quality. Our method ranked second with average values of 0.82, 0.85 and 0.08, exhibiting significantly better ratings than LRP regarding intuitive validity and semantic meaningfulness ($2.81<t(7)<2.88$, $p<.05$). The differences to DeepTaylor were not significant with $0.91<t(7)<1.54$ and $.16<p<.39$. While the inter-observer reliability for intuitive validity and semantic meaningfulness was moderate with $.447$ to $.485$, it was low for the image quality criterion with only $.09$, basically forming two clusters where one preferred parsimonious and the other fully colored annotations (cf.~Sec.~\ref{Discussion}). The $\mathrm{KMO}$ was .646 with the largest prinicipal component accounting for 76.6\,\% of the variance.
This pattern mostly repeated for the trucks-vs.-cars dataset with average z-adjusted values for intuitive validity, semantic meaningfulness and image quality of 2.58, 2.76 and 2.03 for DeepTaylor, and 1.10, 1.55 and 0.77 for our method. While our method achieved significantly higher values than DeepSHAP and LRP for intuitive validity and semantic meaningfulness, no significant differences were observed for image quality. Similarly, the differences to DeepTaylor were not significant with $1.63<t(4)<1.92$. The inter-observer reliability was $\rho=0.612/0.742/0.195$ for intuitive validity, semantic meaningfulness and image quality, respectively, showing a similar pattern as for the cats-vs.-dogs dataset. The $\mathrm{KMO}$ was .736 and the largest principal component explained 88.0\,\% of the variance.
}

\subsection{Discussion}
\label{Discussion}
Based on the results of the user study, our method was able to outperform DeepSHAP, DeepTaylor and LRP in terms of intuitive validity, semantic meaningfulness and image quality on the {\modified{}medical imaging datasets. While for DeepSHAP and LRP the results were significant for all tested criteria, the image quality could not be shown to be significantly different both for the LIDC-IDRI and the BreastMNIST dataset compared to the DeepTaylor algorithm, although the observed effect size was rather high. The results on the CIFAR-10 dataset indicate that our method has its greatest strengths in the visualization of small to medium-sized datasets where a high-quality domain-transfer can successfully be realized using a Cycle-GAN approach. While our method ranked second on CIFAR-10, DeepTaylor was clearly superior in this scenario.} We demonstrated our method to be able to semantically modify {\modified{}medical imaging data} with respect to the classification context. The user study indicates that the visualizations generated by our method emphasize regions which are perceived to be decision-relevant in the given images, while preserving a high image quality as well as intuitive validity.
With an average rank of 1.28, 1.36 and 1.52 {\modified{}on the LIDC-IDRI and 1.75, 1.77 and 1.64 on the BreastMNIST dataset}, our method markedly ranked best in comparison to the analyzed SoA methods {\modified{}on the medical imaging datasets}, followed by DeepTaylor, DeepSHAP and LRP. The extraction of a general factor of preferability $\Phi$, as well as the inter-criteria correlations showed that the measured performance criteria are correlated and influence each other, expressing the need for future work on decision explanation to account for them.
The survey evaluation revealed no significant differences between different questionnaire variants (cf.~Sec.~\ref{Experimental.UserStudy}) and showed a substantial to moderate correlation between the raters {\modified{}for intuitive validity and semantic meaningfulness, as well as for image quality on the medical imaging datasets. As pointed out in Sec.~\ref{Experimental.UserStudy}, the studies on the BreastMNIST and CIFAR datasets were conducted using an online survey. While the participation in the BreastMNIST study was restricted to participants with a significant experience in medical image processing and/or assessment, the CIFAR study did not have any of these restrictions, which allows for a larger variety of interpretations, thus a larger variance and lower inter-observer reliability. Interestingly, this mainly occurred for the image quality, but not for the intuitive validity and semantic meaningfulness criterion.}

\section{Conclusion}
\label{Conclusion}
Deep learning is an important tool for medical image analysis due to its ability to analyze vast amounts of data autonomously. Its low transparency and the resulting lack of understanding of its decision-making process, however, pose a major obstacle for its application in clinical routine. In medical imaging, it is particularly important to understand the inner workings of an algorithm with respect to issues such as algorithm validation, product quality, and finally liability. For this reason, there is a need for methods which allow clinicians as well as engineers to intuitively visualize the network decision process, i.\,e.~to ``see what the network sees.'' As shown in Sec.~\ref{Results}, existing {\modified{}methods can be well-suited for the application on classifiers trained on large datasets of natural images}, but reach this aim only to a limited degree {\modified{}when applied to medical imaging data. For both scenarios there is still room for improvement}. 

With our method we were able to generate high-quality decision explanations for a trained classifier, which are both intuitively valid as well as semantically correct, and could clearly improve on the tested SoA approaches {\modified{}for medical imaging data}. As shown in Sec.~\ref{Results}, our method focuses on decision-relevant areas, and changes in these regions were demonstrated to be associated with changes in the classifier output probabilities.
Regarding the analyzed criteria, it could be shown that intuitive validity, semantic meaningfulness and image quality are closely related, and that therefore algorithms for decision explanation should account for all of them equally. {\modified{}With only 772 and 780 samples, the medical imaging data sets we used were significantly smaller than} comparable computer vision benchmark data sets, such as the CIFAR-10/CIFAR-100 (60,000 samples) \cite{krizhevsky2009learning}, MNIST (70,000 samples) \cite{lecun2010mnist} or cityscapes dataset (20,000 samples) \cite{cordts2016cityscapes}. Nevertheless, our method was able to successfully visualize network decisions, making it a candidate for explaining deep learning-based models for medical image analysis.

With our work, we contribute a significant step towards a better understanding of DNN-based classifier decisions, which in the future could help both engineers as well as clinical practitioners to better understand and hence develop medical algorithms more effectively, which might ultimately lead to an improved overall clinical acceptance. 

\subsection{Limitations}
The quality of our method is determined by the quality of the trained CycleGAN. While recent work suggested approaches for dealing with non-convergence in GANs \cite{berthelot2017began,liu2017approximation}, there is currently no well-established measure of convergence, especially regarding perceptual quality of decision explanations. While in our experiments the model did not appear to behave chaotically, it was not analyzed whether this is a side-effect of the model structure, or was specific to our problem. As our model is based on GAN training, creating a decision explanation network requires a high amount of computational resources, which becomes more relevant with increasing input image sizes. Thus, our approach is applicable to the explanation of fully-trained networks, yet it may be less suitable for immediate visualization and rapid prototyping. Additionally, future work should investigate the applicability to further problems, including tasks other than binary classification.
{\modified{}Notably, our method seems less suited for highly heterogeneous datasets in which realistic domain transfers through a CycleGAN are rather difficult to implement, as was shown in the CIFAR-10 settings. Having large amounts of training data and well-performing classifiers, these setups are the main domains of existing approaches like the DeepTaylor algorithm (or subsequent work, such as Anchors or LORE), as was shown in previous work. Still our method was able to rank second in these scenarios, too. However, our work is specifically tailored to the application in the medical imaging domain, where the compared algorithms did not provide sufficiently intuitive decision explanations. We expect this to be the case due to the sensitivity of these algorithms to either small changes on the pixel-level, or a lack of realism for the local proxy models, as was already pointed out by Apicella, Kindermans and Guidotti \cite{kindermans2019reliability,apicella2020general,guidotti2018local}.}
While we could show significant improvements over the analyzed SoA methods {\modified{} for medical imaging data}, the average raw questionnaire results for our method {\modified{}were in the range between 1.04 and 1.92 on the LIDC-IDRI dataset, and 1.33 and 1.72 for the BreastMNIST dataset} on a scale ranging from -4 to 4, indicating that there is still room for improvement. Compared to -1.52 to -0.85 {\modified{} and .75 to .91 for the second ranked method DeepTaylor, respectively,} our method was able to cover some of this potential. However, it is desirable that future research can further improve on that.

{\modified{}Each of our 34 participants evaluated 24 images for 3 criteria each, resulting in a total of 288 answers per questionnaire and 9{,}792 answers in total. While the number of participants and therefore the degrees of freedom of the used t-test was rather low, due to the large effect sizes statistical significance could still be shown for most comparisons on the medical imaging datasets. Nonetheless, based on the promising results of this study, an even more comprehensive user study in future work would be highly desirable.}

\newpage
\section*{Vitae}

\vspace{2em}

\hspace{-1.9em}
\begin{tabularx}{\textwidth}{lX}
\includegraphics[width=.25\textwidth]{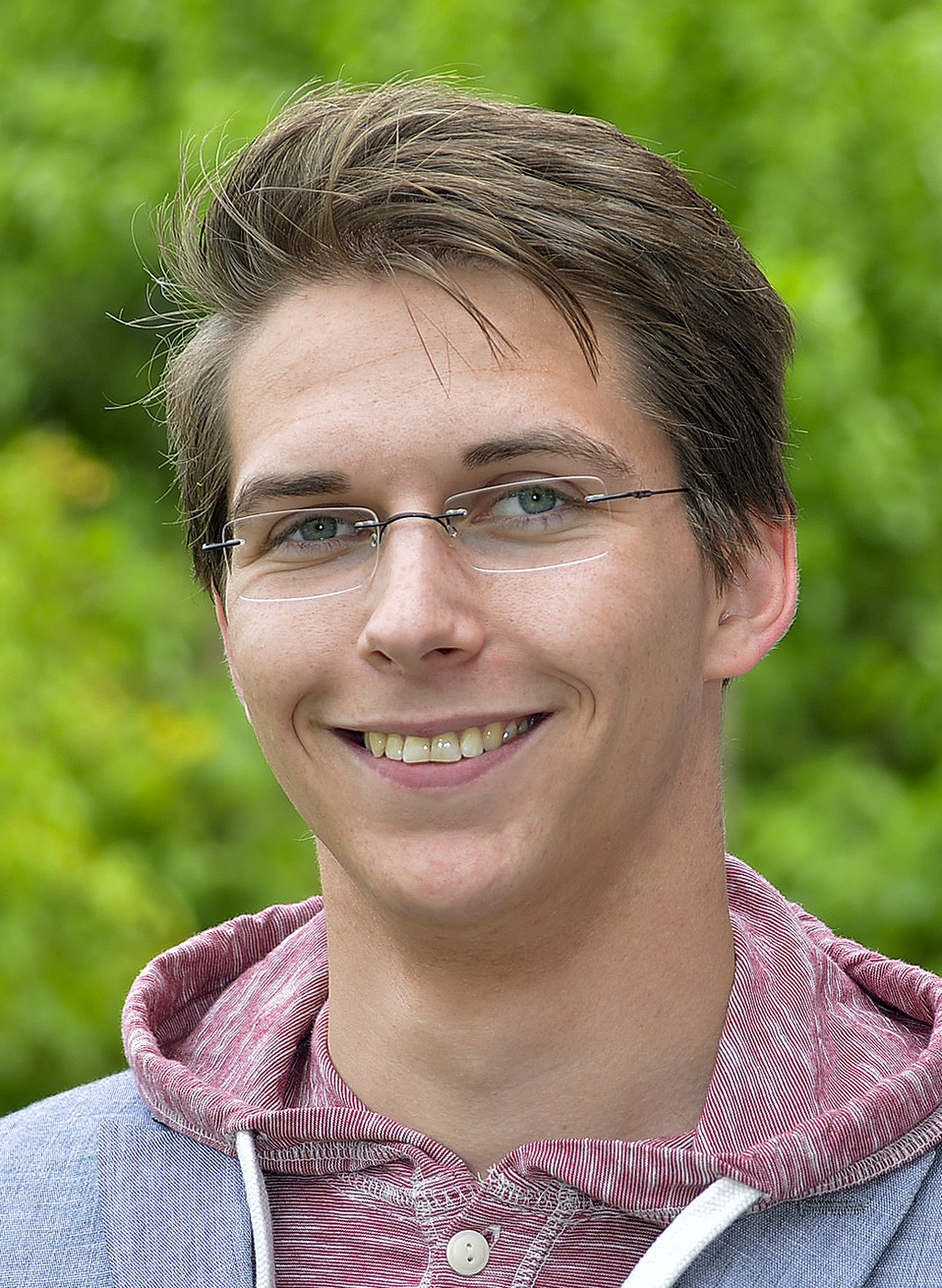}
&\vspace{-.36\textwidth}Alexander Katzmann, is a research scientist at Siemens Healthineers. He holds a M.\,Sc.~degree in computer science from Ilmenau, University of Technology. After researching in the field of neuroinformatics and cognitive robotics, his current work focuses on deep learning-based medical image analysis. He is based in the CT Image Analytics group in Forchheim, Germany, and currently working on his Ph.\,D.~in collaboration with Ilmenau, University of Technology.
\end{tabularx}
\\[.5cm]
\begin{tabularx}{\textwidth}{lX}
\includegraphics[width=.25\textwidth]{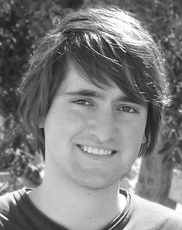}
&\vspace{-.33\textwidth}Dr.~Oliver Taubmann is a research scientist with Siemens Healthineers. As a member of the CT Image Analytics group based in Forchheim, Germany, he develops clinical prototypes for learning-based image analysis to better support radiologists. Oliver holds a Ph.\,D.~degree in computer science from FAU Erlangen-N{\"u}rnberg, for which he devised novel techniques for time-resolved cardiac image reconstruction from rotational angiography in the interventional suite.
\end{tabularx}
\\[.5cm]
\begin{tabularx}{\textwidth}{lX}
\includegraphics[width=.25\textwidth]{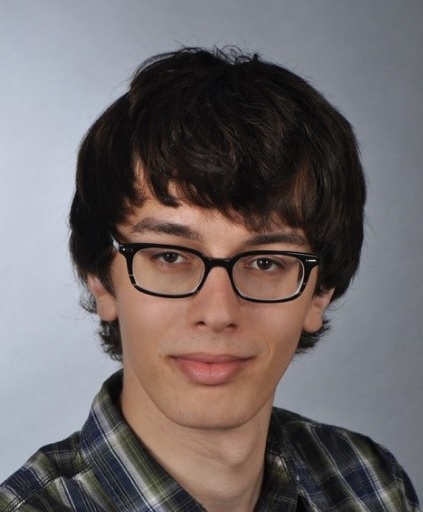}
&\vspace{-.33\textwidth}Stephen Ahmad is a software developer at Siemens Healthineers. He holds a M.\,Sc.~degree in engineering informatics and is currently based in the department of Digital Health Imaging Decision Support, Erlangen, Germany. He is focused on AI-related algorithm engineering and currently a developer for the Siemens Healthineers ``AI-Rad Companion''. 
\end{tabularx}
\\[.5cm]
\\[.5cm]
\begin{tabularx}{\textwidth}{lX}
\includegraphics[width=.25\textwidth]{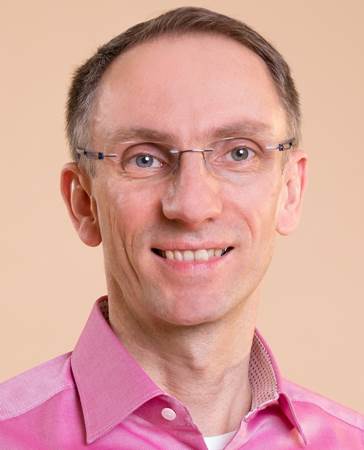}
&\vspace{-.328\textwidth}Dr.~Michael S\"uhling is head of the CT Image Analytics group at the Siemens Healthineers CT headquarter in Forchheim, Germany. His research interests are mainly in the area of image analysis technologies for CT scan automation and clinical decision making support. He has been reviewer for the IEEE Transactions on Medical Imaging and the IEEE Transactions on Image Processing.
\end{tabularx}
\\[.5cm]
\begin{tabularx}{\textwidth}{lX}
\includegraphics[width=.25\textwidth]{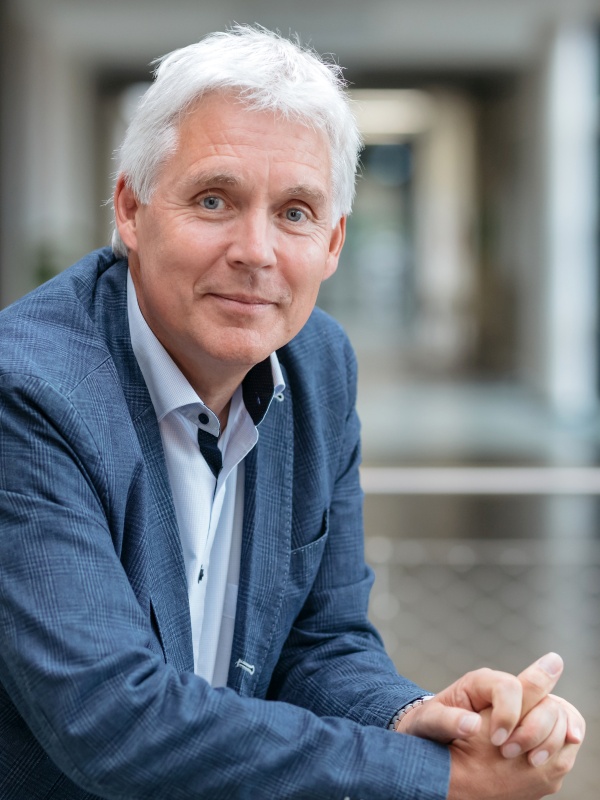}
&\vspace{-.35\textwidth}Horst-Michael Gro\ss~is full professor for Computer Science at Ilmenau University of Technology and head of the research lab (chair) for Neuroinformatics and Cognitive Robotics. From methodological view, his research is focused on real-time approaches to person detection, tracking, and re-identification in real-world video data streams and deep learning approaches for image-based object recognition, semantic segmentation, pose recognition, grasp pose estimation, and action recognition. Amongst others, he is a member of the IEEE and the International Neural Networks Society (INNS).
\end{tabularx}

\section*{Acknowledgements}
This work has received funding from the German Federal Ministry of Education and Research as part of the PANTHER project under grant agreement no.~13GW0163A.

\section*{Disclaimer}
The concepts and information presented in this article are based on research and are not commercially available.

\bibliography{elsarticle-template}

\newpage
\section*{Appendix}
\paragraph{Questionnaires}
{\modified{}The following pages exemplarily cover the questionnaires (A/B) of the LIDC-IDRI user study described in Sec.~\ref{Experimental.UserStudy}}. Each user received a study introduction sheet (first page), followed by 8 pages including 12 benign and 12 malignant lesions. {\modified{}The questionnaires for the other studies were analogously structured and were conducted using web-based surveys.} 


\section*{Supplementary material}

## Deep Decision Visualization for Lung Malignancy Estimation
### User Study

Deep learning has become the dominating method in artificial intelligence (**AI**). While the use of deep learning goes hand in hand with major benefits regarding the overall classification accuracy, it generally suffers from being difficult to explain, and to comprehend. Deep learning is often seen as a black box, which poses a major obstacle for the daily clinical use. While deep learning seems able to do such a classification, its low explainability leads to clinicians being reluctant of employing the method.

Lung lesion malignancy is an example for a well gradable problem when using deep learning, as demonstrated in recent studies[1,2]. Typical visual criteria[3] are depicted in the following:

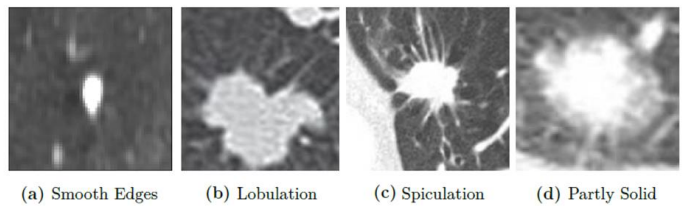

(a) Smooth Edges    (b) Lobulation    (c) Spiculation    (d) Partly Solid

With **a)** smooth edges typically indicating benign lesions, **b)** lobulation indicating an intermediate probability for malignancy, and **c)** spiculation and **d)** partly solidness being clear indicators for malignancy. Note that lesion margins are most important for malignancy classification in deep neural networks, as they are clinically, too.

In recent years, decision visualization algorithms have been proposed to project network decisions back to the input space by visualizing **decision-relevant regions**. Due to the network structure, however, this projection can become ambiguous, allowing for multiple explanations. For clinical use, it is therefore important for the explanations to be a) most intuitive, as well as b) semantically meaningful. Furthermore, c) decision visualizations should provide the highest possible quality.

Within our user study, we analyze multiple methods for visualizing deep learning decisions by projecting them back to the original image input. In all images, Input regions are colored 1.) **blue for benign**, and 2.) **red for regions indicating malignancy**. Your task is to evaluate all variants in terms of **a) intuitivity**, **b) semantical meaningfulness**, according to the above criteria, and **c) image quality**.

For each of the given images, please evaluate each criterion in a scale ranging from -4 (very low) to +4 (very high).

1

**Benign lesions (A)**

All of the following depicted lesions are benign, as given by the ground truth data. Also, all lesions are recognized as such by the classifier, i.e. the images contain **no false predictions**. For each image, **benignity is indicated in blue** while **malignancy is indicated in red**. Please rate **a)** intuitivity, **b)** semantical meaningfulness, and **c)** image quality. For each criterion, please use a scale from **-4 (very low) to 4 (very high)**.

1)

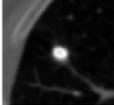 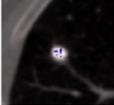 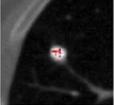 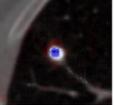 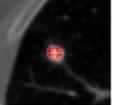

```
                    -   |0|   +          -   |0|   +          -   |0|   +          -   |0|   +
Intuitivity:      OOOO|O|OOOO          OOOO|O|OOOO          OOOO|O|OOOO          OOOO|O|OOOO
Semantic:         OOOO|O|OOOO          OOOO|O|OOOO          OOOO|O|OOOO          OOOO|O|OOOO
Image Quality:    OOOO|O|OOOO          OOOO|O|OOOO          OOOO|O|OOOO          OOOO|O|OOOO
```

2)

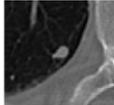 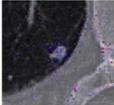 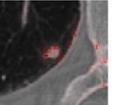 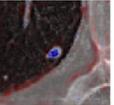 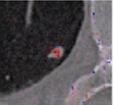

```
                    -   |0|   +          -   |0|   +          -   |0|   +          -   |0|   +
Intuitivity:      OOOO|O|OOOO          OOOO|O|OOOO          OOOO|O|OOOO          OOOO|O|OOOO
Semantic:         OOOO|O|OOOO          OOOO|O|OOOO          OOOO|O|OOOO          OOOO|O|OOOO
Image Quality:    OOOO|O|OOOO          OOOO|O|OOOO          OOOO|O|OOOO          OOOO|O|OOOO
```

3)

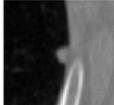 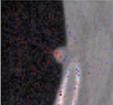 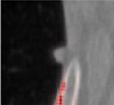 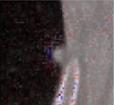 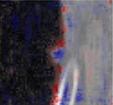

```
                    -   |0|   +          -   |0|   +          -   |0|   +          -   |0|   +
Intuitivity:      OOOO|O|OOOO          OOOO|O|OOOO          OOOO|O|OOOO          OOOO|O|OOOO
Semantic:         OOOO|O|OOOO          OOOO|O|OOOO          OOOO|O|OOOO          OOOO|O|OOOO
Image Quality:    OOOO|O|OOOO          OOOO|O|OOOO          OOOO|O|OOOO          OOOO|O|OOOO
```

AB4)

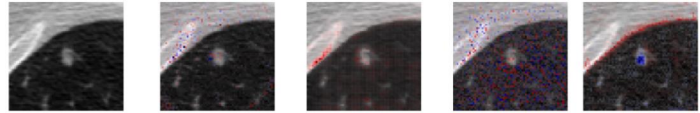

```
                     -   |0|   +       -   |0|   +       -   |0|   +       -   |0|   +
Intuitivity:      0000|0|0000       0000|0|0000       0000|0|0000       0000|0|0000
Semantic:         0000|0|0000       0000|0|0000       0000|0|0000       0000|0|0000
Image Quality:    0000|0|0000       0000|0|0000       0000|0|0000       0000|0|0000
```

5)

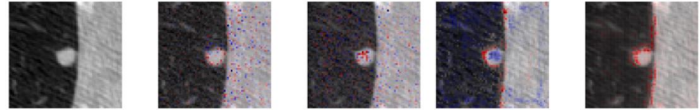

```
                     -   |0|   +       -   |0|   +       -   |0|   +       -   |0|   +
Intuitivity:      0000|0|0000       0000|0|0000       0000|0|0000       0000|0|0000
Semantic:         0000|0|0000       0000|0|0000       0000|0|0000       0000|0|0000
Image Quality:    0000|0|0000       0000|0|0000       0000|0|0000       0000|0|0000
```

6)

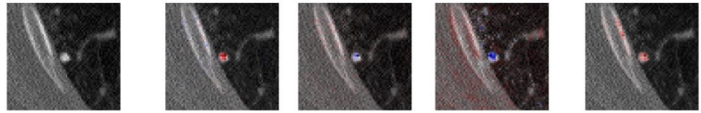

```
                     -   |0|   +       -   |0|   +       -   |0|   +       -   |0|   +
Intuitivity:      0000|0|0000       0000|0|0000       0000|0|0000       0000|0|0000
Semantic:         0000|0|0000       0000|0|0000       0000|0|0000       0000|0|0000
Image Quality:    0000|0|0000       0000|0|0000       0000|0|0000       0000|0|0000
```

7)

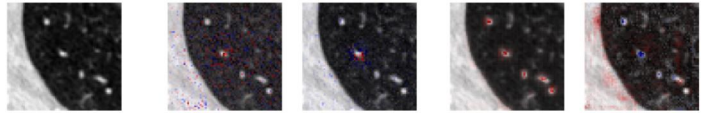

```
                     -   |0|   +       -   |0|   +       -   |0|   +       -   |0|   +
Intuitivity:      0000|0|0000       0000|0|0000       0000|0|0000       0000|0|0000
Semantic:         0000|0|0000       0000|0|0000       0000|0|0000       0000|0|0000
Image Quality:    0000|0|0000       0000|0|0000       0000|0|0000       0000|0|0000
```

3

AB8)

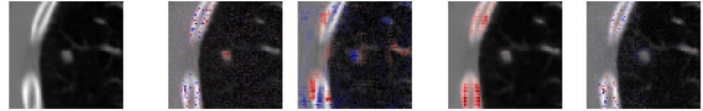

```
                    -  |0|  +      -  |0|  +      -  |0|  +      -  |0|  +
Intuitivity:     0000|0|0000   0000|0|0000   0000|0|0000   0000|0|0000
Semantic:        0000|0|0000   0000|0|0000   0000|0|0000   0000|0|0000
Image Quality:   0000|0|0000   0000|0|0000   0000|0|0000   0000|0|0000
```

9)

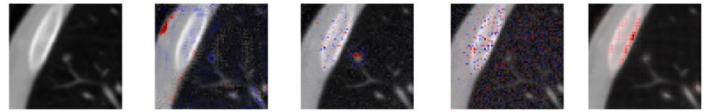

```
                    -  |0|  +      -  |0|  +      -  |0|  +      -  |0|  +
Intuitivity:     0000|0|0000   0000|0|0000   0000|0|0000   0000|0|0000
Semantic:        0000|0|0000   0000|0|0000   0000|0|0000   0000|0|0000
Image Quality:   0000|0|0000   0000|0|0000   0000|0|0000   0000|0|0000
```

10)

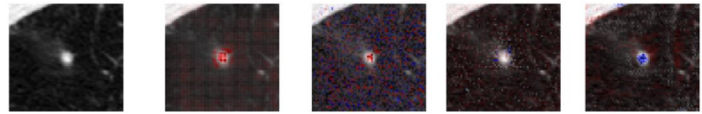

```
                    -  |0|  +      -  |0|  +      -  |0|  +      -  |0|  +
Intuitivity:     0000|0|0000   0000|0|0000   0000|0|0000   0000|0|0000
Semantic:        0000|0|0000   0000|0|0000   0000|0|0000   0000|0|0000
Image Quality:   0000|0|0000   0000|0|0000   0000|0|0000   0000|0|0000
```

11)

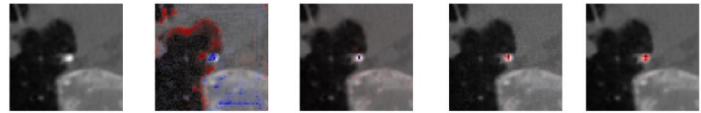

```
                    -  |0|  +      -  |0|  +      -  |0|  +      -  |0|  +
Intuitivity:     0000|0|0000   0000|0|0000   0000|0|0000   0000|0|0000
Semantic:        0000|0|0000   0000|0|0000   0000|0|0000   0000|0|0000
Image Quality:   0000|0|0000   0000|0|0000   0000|0|0000   0000|0|0000
```

4

54Published article – DOI: 10.1016/j.neucom.2021.05.081

AB12)

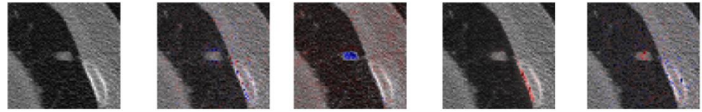

```
                     -   |0|   +       -   |0|   +       -   |0|   +       -   |0|   +
Intuitivity:      OOOO|O|OOOO        OOOO|O|OOOO        OOOO|O|OOOO        OOOO|O|OOOO
Semantic:         OOOO|O|OOOO        OOOO|O|OOOO        OOOO|O|OOOO        OOOO|O|OOOO
Image Quality:    OOOO|O|OOOO        OOOO|O|OOOO        OOOO|O|OOOO        OOOO|O|OOOO
```

5

**Malignant lesions (A)**

All of the following depicted lesions are malignant, as given by the ground truth data. Also, all lesions are recognized as such by the classifier, i.e. the images contain **no false predictions**. For each image, **benignity is indicated in blue** while **malignancy is indicated in red**. Please rate **a)** intuitivity, **b)** semantical meaningfulness, and **c)** image quality. For each criterion, please use a scale from **-4 (very low) to 4 (very high)**.

1.)

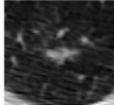 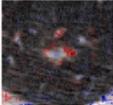 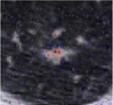 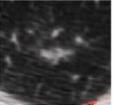 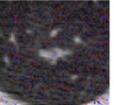

```
                         -   |0|   +      -   |0|   +      -   |0|   +      -   |0|   +
Intuitivity:           OOOO|O|OOOO      OOOO|O|OOOO      OOOO|O|OOOO      OOOO|O|OOOO
Semantic:              OOOO|O|OOOO      OOOO|O|OOOO      OOOO|O|OOOO      OOOO|O|OOOO
Image Quality:         OOOO|O|OOOO      OOOO|O|OOOO      OOOO|O|OOOO      OOOO|O|OOOO
```

2)

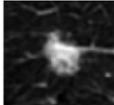 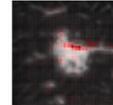 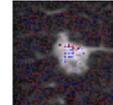 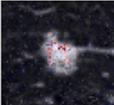 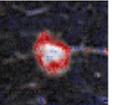

```
                         -   |0|   +      -   |0|   +      -   |0|   +      -   |0|   +
Intuitivity:           OOOO|O|OOOO      OOOO|O|OOOO      OOOO|O|OOOO      OOOO|O|OOOO
Semantic:              OOOO|O|OOOO      OOOO|O|OOOO      OOOO|O|OOOO      OOOO|O|OOOO
Image Quality:         OOOO|O|OOOO      OOOO|O|OOOO      OOOO|O|OOOO      OOOO|O|OOOO
```

3)

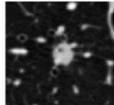 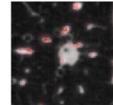 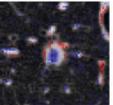 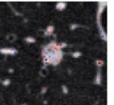 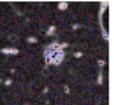

```
                         -   |0|   +      -   |0|   +      -   |0|   +      -   |0|   +
Intuitivity:           OOOO|O|OOOO      OOOO|O|OOOO      OOOO|O|OOOO      OOOO|O|OOOO
Semantic:              OOOO|O|OOOO      OOOO|O|OOOO      OOOO|O|OOOO      OOOO|O|OOOO
Image Quality:         OOOO|O|OOOO      OOOO|O|OOOO      OOOO|O|OOOO      OOOO|O|OOOO
```

6

AM4)

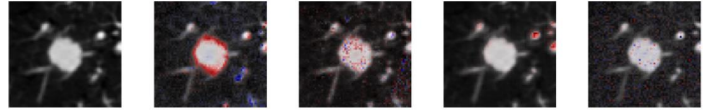

```
                  -  |0|  +       -  |0|  +       -  |0|  +       -  |0|  +
Intuitivity:    OOOO|O|OOOO     OOOO|O|OOOO     OOOO|O|OOOO     OOOO|O|OOOO
Semantic:       OOOO|O|OOOO     OOOO|O|OOOO     OOOO|O|OOOO     OOOO|O|OOOO
Image Quality:  OOOO|O|OOOO     OOOO|O|OOOO     OOOO|O|OOOO     OOOO|O|OOOO
```

5)

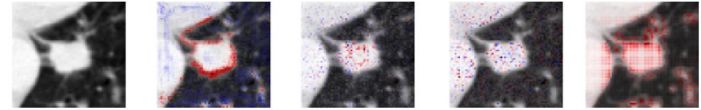

```
                  -  |0|  +       -  |0|  +       -  |0|  +       -  |0|  +
Intuitivity:    OOOO|O|OOOO     OOOO|O|OOOO     OOOO|O|OOOO     OOOO|O|OOOO
Semantic:       OOOO|O|OOOO     OOOO|O|OOOO     OOOO|O|OOOO     OOOO|O|OOOO
Image Quality:  OOOO|O|OOOO     OOOO|O|OOOO     OOOO|O|OOOO     OOOO|O|OOOO
```

6)

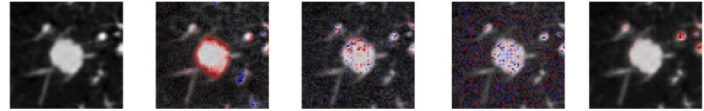

```
                  -  |0|  +       -  |0|  +       -  |0|  +       -  |0|  +
Intuitivity:    OOOO|O|OOOO     OOOO|O|OOOO     OOOO|O|OOOO     OOOO|O|OOOO
Semantic:       OOOO|O|OOOO     OOOO|O|OOOO     OOOO|O|OOOO     OOOO|O|OOOO
Image Quality:  OOOO|O|OOOO     OOOO|O|OOOO     OOOO|O|OOOO     OOOO|O|OOOO
```

7)

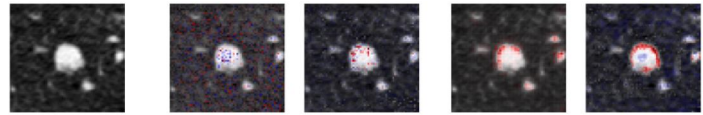

```
                  -  |0|  +       -  |0|  +       -  |0|  +       -  |0|  +
Intuitivity:    OOOO|O|OOOO     OOOO|O|OOOO     OOOO|O|OOOO     OOOO|O|OOOO
Semantic:       OOOO|O|OOOO     OOOO|O|OOOO     OOOO|O|OOOO     OOOO|O|OOOO
Image Quality:  OOOO|O|OOOO     OOOO|O|OOOO     OOOO|O|OOOO     OOOO|O|OOOO
```

7

**AM8)**

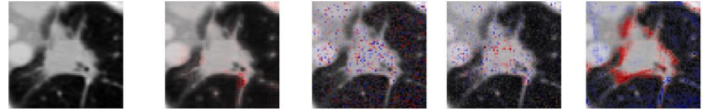

|  | - \|0\| + | - \|0\| + | - \|0\| + | - \|0\| + |
|---|---|---|---|---|
| Intuitivity: | OOOO\|O\|OOOO | OOOO\|O\|OOOO | OOOO\|O\|OOOO | OOOO\|O\|OOOO |
| Semantic: | OOOO\|O\|OOOO | OOOO\|O\|OOOO | OOOO\|O\|OOOO | OOOO\|O\|OOOO |
| Image Quality: | OOOO\|O\|OOOO | OOOO\|O\|OOOO | OOOO\|O\|OOOO | OOOO\|O\|OOOO |

**9)**

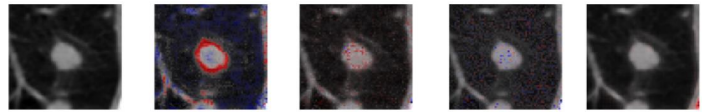

|  | - \|0\| + | - \|0\| + | - \|0\| + | - \|0\| + |
|---|---|---|---|---|
| Intuitivity: | OOOO\|O\|OOOO | OOOO\|O\|OOOO | OOOO\|O\|OOOO | OOOO\|O\|OOOO |
| Semantic: | OOOO\|O\|OOOO | OOOO\|O\|OOOO | OOOO\|O\|OOOO | OOOO\|O\|OOOO |
| Image Quality: | OOOO\|O\|OOOO | OOOO\|O\|OOOO | OOOO\|O\|OOOO | OOOO\|O\|OOOO |

**10)**

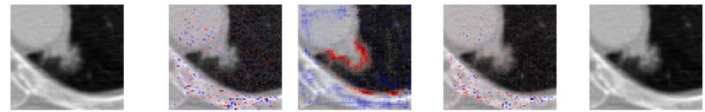

|  | - \|0\| + | - \|0\| + | - \|0\| + | - \|0\| + |
|---|---|---|---|---|
| Intuitivity: | OOOO\|O\|OOOO | OOOO\|O\|OOOO | OOOO\|O\|OOOO | OOOO\|O\|OOOO |
| Semantic: | OOOO\|O\|OOOO | OOOO\|O\|OOOO | OOOO\|O\|OOOO | OOOO\|O\|OOOO |
| Image Quality: | OOOO\|O\|OOOO | OOOO\|O\|OOOO | OOOO\|O\|OOOO | OOOO\|O\|OOOO |

**11)**

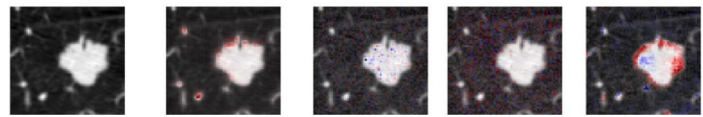

|  | - \|0\| + | - \|0\| + | - \|0\| + | - \|0\| + |
|---|---|---|---|---|
| Intuitivity: | OOOO\|O\|OOOO | OOOO\|O\|OOOO | OOOO\|O\|OOOO | OOOO\|O\|OOOO |
| Semantic: | OOOO\|O\|OOOO | OOOO\|O\|OOOO | OOOO\|O\|OOOO | OOOO\|O\|OOOO |
| Image Quality: | OOOO\|O\|OOOO | OOOO\|O\|OOOO | OOOO\|O\|OOOO | OOOO\|O\|OOOO |

8

**AM12)**

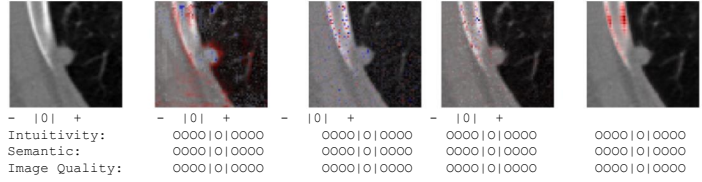

```
                -   |0|   +      -   |0|   +      -   |0|   +      -   |0|   +          -   |0|   +
Intuitivity:     0000|0|0000      0000|0|0000      0000|0|0000      0000|0|0000          0000|0|0000
Semantic:        0000|0|0000      0000|0|0000      0000|0|0000      0000|0|0000          0000|0|0000
Image Quality:   0000|0|0000      0000|0|0000      0000|0|0000      0000|0|0000          0000|0|0000
```

9

59

**Benign lesions (B)**

All of the following depicted lesions are benign, as given by the ground truth data. Also, all lesions are recognized as such by the classifier, i.e. the images contain **no false predictions**. For each image, **benignity is indicated in blue** while **malignancy is indicated in red**. Please rate **a)** intuitivity, **b)** semantical meaningfulness, and **c)** image quality. For each criterion, please use a scale from **-4 (very low) to 4 (very high)**.

1)

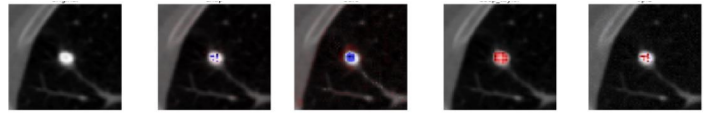

```
                    -   |0|   +         -   |0|   +         -   |0|   +         -   |0|   +
Intuitivity:      OOOO|O|OOOO         OOOO|O|OOOO         OOOO|O|OOOO         OOOO|O|OOOO
Semantic:         OOOO|O|OOOO         OOOO|O|OOOO         OOOO|O|OOOO         OOOO|O|OOOO
Image Quality:    OOOO|O|OOOO         OOOO|O|OOOO         OOOO|O|OOOO         OOOO|O|OOOO
```

2)

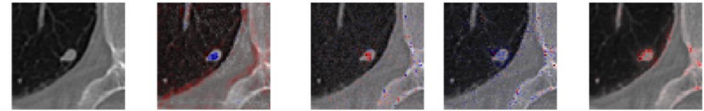

```
                    -   |0|   +         -   |0|   +         -   |0|   +         -   |0|   +
Intuitivity:      OOOO|O|OOOO         OOOO|O|OOOO         OOOO|O|OOOO         OOOO|O|OOOO
Semantic:         OOOO|O|OOOO         OOOO|O|OOOO         OOOO|O|OOOO         OOOO|O|OOOO
Image Quality:    OOOO|O|OOOO         OOOO|O|OOOO         OOOO|O|OOOO         OOOO|O|OOOO
```

3)

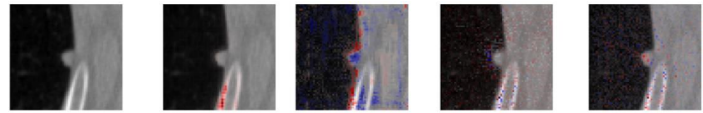

```
                    -   |0|   +         -   |0|   +         -   |0|   +         -   |0|   +
Intuitivity:      OOOO|O|OOOO         OOOO|O|OOOO         OOOO|O|OOOO         OOOO|O|OOOO
Semantic:         OOOO|O|OOOO         OOOO|O|OOOO         OOOO|O|OOOO         OOOO|O|OOOO
Image Quality:    OOOO|O|OOOO         OOOO|O|OOOO         OOOO|O|OOOO         OOOO|O|OOOO
```

2

BB4)

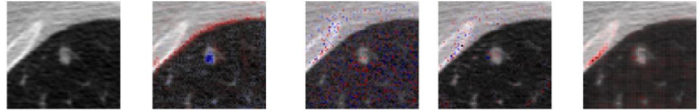

```
                     -  |0|  +      -  |0|  +      -  |0|  +      -  |0|  +
Intuitivity:      OOOO|O|OOOO   OOOO|O|OOOO   OOOO|O|OOOO   OOOO|O|OOOO
Semantic:         OOOO|O|OOOO   OOOO|O|OOOO   OOOO|O|OOOO   OOOO|O|OOOO
Image Quality:    OOOO|O|OOOO   OOOO|O|OOOO   OOOO|O|OOOO   OOOO|O|OOOO
```

5)

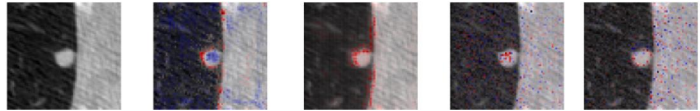

```
                     -  |0|  +      -  |0|  +      -  |0|  +      -  |0|  +
Intuitivity:      OOOO|O|OOOO   OOOO|O|OOOO   OOOO|O|OOOO   OOOO|O|OOOO
Semantic:         OOOO|O|OOOO   OOOO|O|OOOO   OOOO|O|OOOO   OOOO|O|OOOO
Image Quality:    OOOO|O|OOOO   OOOO|O|OOOO   OOOO|O|OOOO   OOOO|O|OOOO
```

6)

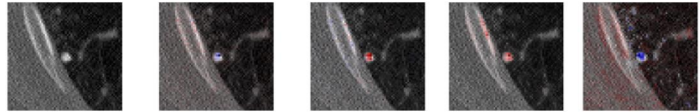

```
                     -  |0|  +      -  |0|  +      -  |0|  +      -  |0|  +
Intuitivity:      OOOO|O|OOOO   OOOO|O|OOOO   OOOO|O|OOOO   OOOO|O|OOOO
Semantic:         OOOO|O|OOOO   OOOO|O|OOOO   OOOO|O|OOOO   OOOO|O|OOOO
Image Quality:    OOOO|O|OOOO   OOOO|O|OOOO   OOOO|O|OOOO   OOOO|O|OOOO
```

7)

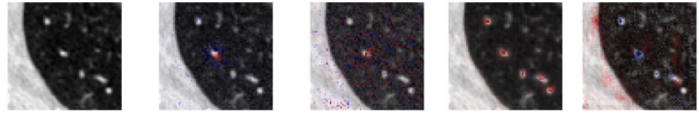

```
                     -  |0|  +      -  |0|  +      -  |0|  +      -  |0|  +
Intuitivity:      OOOO|O|OOOO   OOOO|O|OOOO   OOOO|O|OOOO   OOOO|O|OOOO
Semantic:         OOOO|O|OOOO   OOOO|O|OOOO   OOOO|O|OOOO   OOOO|O|OOOO
Image Quality:    OOOO|O|OOOO   OOOO|O|OOOO   OOOO|O|OOOO   OOOO|O|OOOO
```

3

BB8)

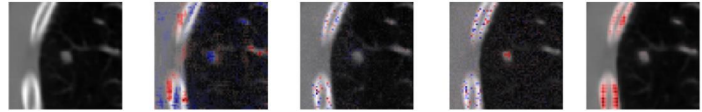

```
                     -  |0|  +       -  |0|  +       -  |0|  +       -  |0|  +
Intuitivity:       OOOO|O|OOOO     OOOO|O|OOOO     OOOO|O|OOOO     OOOO|O|OOOO
Semantic:          OOOO|O|OOOO     OOOO|O|OOOO     OOOO|O|OOOO     OOOO|O|OOOO
Image Quality:     OOOO|O|OOOO     OOOO|O|OOOO     OOOO|O|OOOO     OOOO|O|OOOO
```

9)

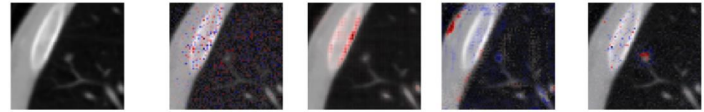

```
                     -  |0|  +       -  |0|  +       -  |0|  +       -  |0|  +
Intuitivity:       OOOO|O|OOOO     OOOO|O|OOOO     OOOO|O|OOOO     OOOO|O|OOOO
Semantic:          OOOO|O|OOOO     OOOO|O|OOOO     OOOO|O|OOOO     OOOO|O|OOOO
Image Quality:     OOOO|O|OOOO     OOOO|O|OOOO     OOOO|O|OOOO     OOOO|O|OOOO
```

10)

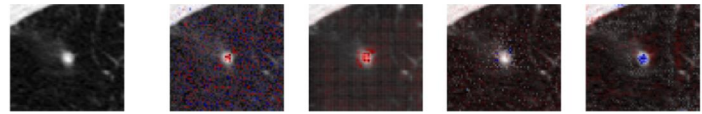

```
                     -  |0|  +       -  |0|  +       -  |0|  +       -  |0|  +
Intuitivity:       OOOO|O|OOOO     OOOO|O|OOOO     OOOO|O|OOOO     OOOO|O|OOOO
Semantic:          OOOO|O|OOOO     OOOO|O|OOOO     OOOO|O|OOOO     OOOO|O|OOOO
Image Quality:     OOOO|O|OOOO     OOOO|O|OOOO     OOOO|O|OOOO     OOOO|O|OOOO
```

11)

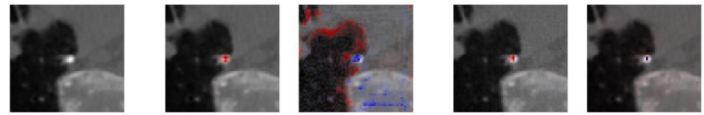

```
                     -  |0|  +       -  |0|  +       -  |0|  +       -  |0|  +
Intuitivity:       OOOO|O|OOOO     OOOO|O|OOOO     OOOO|O|OOOO     OOOO|O|OOOO
Semantic:          OOOO|O|OOOO     OOOO|O|OOOO     OOOO|O|OOOO     OOOO|O|OOOO
Image Quality:     OOOO|O|OOOO     OOOO|O|OOOO     OOOO|O|OOOO     OOOO|O|OOOO
```

4

BB12)

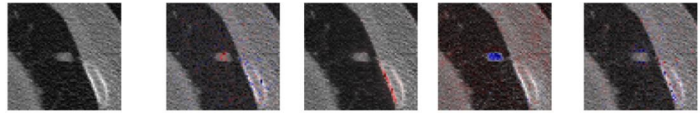

|                | -  |0|  +     | -  |0|  +     | -  |0|  +     | -  |0|  +     |
|----------------|----------------|----------------|----------------|----------------|
| Intuitivity:   | OOOO|O|OOOO    | OOOO|O|OOOO    | OOOO|O|OOOO    | OOOO|O|OOOO    |
| Semantic:      | OOOO|O|OOOO    | OOOO|O|OOOO    | OOOO|O|OOOO    | OOOO|O|OOOO    |
| Image Quality: | OOOO|O|OOOO    | OOOO|O|OOOO    | OOOO|O|OOOO    | OOOO|O|OOOO    |

5

63

**Malignant lesions (B)**

All of the following depicted lesions are malignant, as given by the ground truth data. Also, all lesions are recognized as such by the classifier, i.e. the images contain **no false predictions**. For each image, **benignity is indicated in blue** while **malignancy is indicated in red**. Please rate **a)** intuitivity, **b)** semantical meaningfulness, and **c)** image quality. For each criterion, please use a scale from **-4 (very low) to 4 (very high)**.

1)
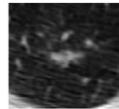 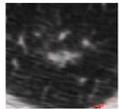 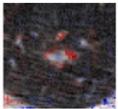 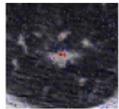 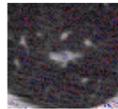

```
                    -   |0|   +        -   |0|   +        -   |0|   +        -   |0|   +
Intuitivity:      OOOO|O|OOOO       OOOO|O|OOOO       OOOO|O|OOOO       OOOO|O|OOOO
Semantic:         OOOO|O|OOOO       OOOO|O|OOOO       OOOO|O|OOOO       OOOO|O|OOOO
Image Quality:    OOOO|O|OOOO       OOOO|O|OOOO       OOOO|O|OOOO       OOOO|O|OOOO
```

2)
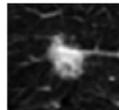 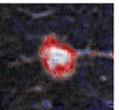 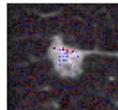 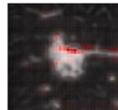 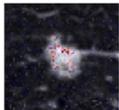

```
                    -   |0|   +        -   |0|   +        -   |0|   +        -   |0|   +
Intuitivity:      OOOO|O|OOOO       OOOO|O|OOOO       OOOO|O|OOOO       OOOO|O|OOOO
Semantic:         OOOO|O|OOOO       OOOO|O|OOOO       OOOO|O|OOOO       OOOO|O|OOOO
Image Quality:    OOOO|O|OOOO       OOOO|O|OOOO       OOOO|O|OOOO       OOOO|O|OOOO
```

3)
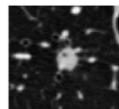 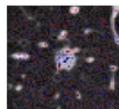 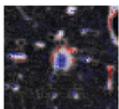 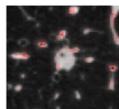 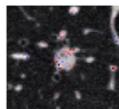

```
                    -   |0|   +        -   |0|   +        -   |0|   +        -   |0|   +
Intuitivity:      OOOO|O|OOOO       OOOO|O|OOOO       OOOO|O|OOOO       OOOO|O|OOOO
Semantic:         OOOO|O|OOOO       OOOO|O|OOOO       OOOO|O|OOOO       OOOO|O|OOOO
Image Quality:    OOOO|O|OOOO       OOOO|O|OOOO       OOOO|O|OOOO       OOOO|O|OOOO
```

6

BM4)

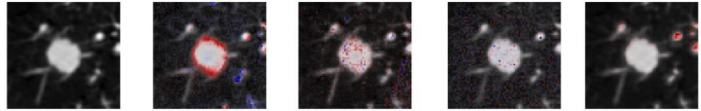

```
                   -  |0|  +      -  |0|  +      -  |0|  +      -  |0|  +
Intuitivity:    OOOO|O|OOOO    OOOO|O|OOOO    OOOO|O|OOOO    OOOO|O|OOOO
Semantic:       OOOO|O|OOOO    OOOO|O|OOOO    OOOO|O|OOOO    OOOO|O|OOOO
Image Quality:  OOOO|O|OOOO    OOOO|O|OOOO    OOOO|O|OOOO    OOOO|O|OOOO
```

5)

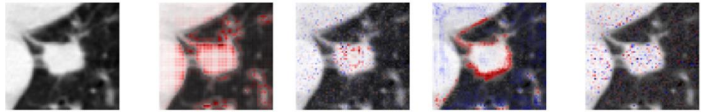

```
                   -  |0|  +      -  |0|  +      -  |0|  +      -  |0|  +
Intuitivity:    OOOO|O|OOOO    OOOO|O|OOOO    OOOO|O|OOOO    OOOO|O|OOOO
Semantic:       OOOO|O|OOOO    OOOO|O|OOOO    OOOO|O|OOOO    OOOO|O|OOOO
Image Quality:  OOOO|O|OOOO    OOOO|O|OOOO    OOOO|O|OOOO    OOOO|O|OOOO
```

6)

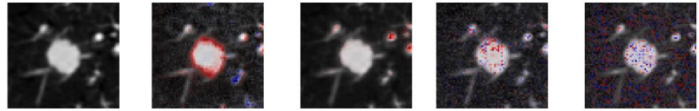

```
                   -  |0|  +      -  |0|  +      -  |0|  +      -  |0|  +
Intuitivity:    OOOO|O|OOOO    OOOO|O|OOOO    OOOO|O|OOOO    OOOO|O|OOOO
Semantic:       OOOO|O|OOOO    OOOO|O|OOOO    OOOO|O|OOOO    OOOO|O|OOOO
Image Quality:  OOOO|O|OOOO    OOOO|O|OOOO    OOOO|O|OOOO    OOOO|O|OOOO
```

7)

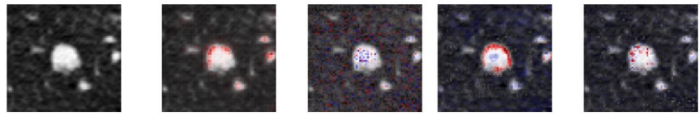

```
                   -  |0|  +      -  |0|  +      -  |0|  +      -  |0|  +
Intuitivity:    OOOO|O|OOOO    OOOO|O|OOOO    OOOO|O|OOOO    OOOO|O|OOOO
Semantic:       OOOO|O|OOOO    OOOO|O|OOOO    OOOO|O|OOOO    OOOO|O|OOOO
Image Quality:  OOOO|O|OOOO    OOOO|O|OOOO    OOOO|O|OOOO    OOOO|O|OOOO
```

7

BM8)

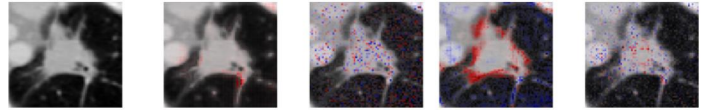

```
                -  |0|  +      -  |0|  +      -  |0|  +      -  |0|  +
Intuitivity:    OOOO|O|OOOO    OOOO|O|OOOO    OOOO|O|OOOO    OOOO|O|OOOO
Semantic:       OOOO|O|OOOO    OOOO|O|OOOO    OOOO|O|OOOO    OOOO|O|OOOO
Image Quality:  OOOO|O|OOOO    OOOO|O|OOOO    OOOO|O|OOOO    OOOO|O|OOOO
```

9)

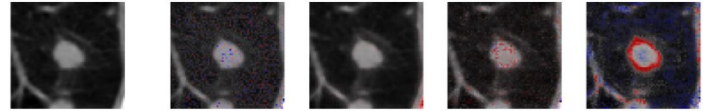

```
                -  |0|  +      -  |0|  +      -  |0|  +      -  |0|  +
Intuitivity:    OOOO|O|OOOO    OOOO|O|OOOO    OOOO|O|OOOO    OOOO|O|OOOO
Semantic:       OOOO|O|OOOO    OOOO|O|OOOO    OOOO|O|OOOO    OOOO|O|OOOO
Image Quality:  OOOO|O|OOOO    OOOO|O|OOOO    OOOO|O|OOOO    OOOO|O|OOOO
```

10)

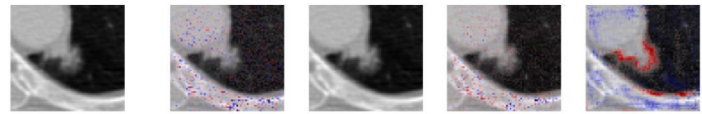

```
                -  |0|  +      -  |0|  +      -  |0|  +      -  |0|  +
Intuitivity:    OOOO|O|OOOO    OOOO|O|OOOO    OOOO|O|OOOO    OOOO|O|OOOO
Semantic:       OOOO|O|OOOO    OOOO|O|OOOO    OOOO|O|OOOO    OOOO|O|OOOO
Image Quality:  OOOO|O|OOOO    OOOO|O|OOOO    OOOO|O|OOOO    OOOO|O|OOOO
```

11)

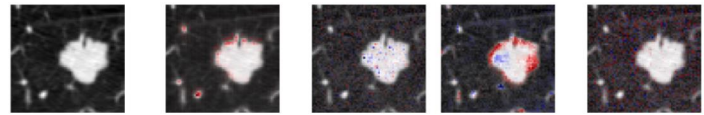

```
                -  |0|  +      -  |0|  +      -  |0|  +      -  |0|  +
Intuitivity:    OOOO|O|OOOO    OOOO|O|OOOO    OOOO|O|OOOO    OOOO|O|OOOO
Semantic:       OOOO|O|OOOO    OOOO|O|OOOO    OOOO|O|OOOO    OOOO|O|OOOO
Image Quality:  OOOO|O|OOOO    OOOO|O|OOOO    OOOO|O|OOOO    OOOO|O|OOOO
```

8

BM12)

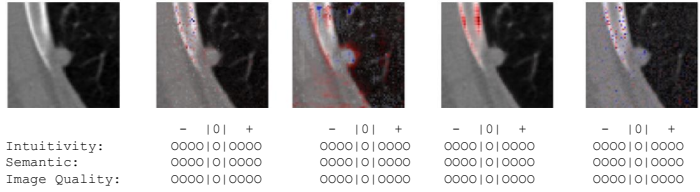

|                | −  \|0\|  +     | −  \|0\|  +     | −  \|0\|  +     | −  \|0\|  +     |
|----------------|-----------------|-----------------|-----------------|-----------------|
| Intuitivity:   | OOOO\|O\|OOOO   | OOOO\|O\|OOOO   | OOOO\|O\|OOOO   | OOOO\|O\|OOOO   |
| Semantic:      | OOOO\|O\|OOOO   | OOOO\|O\|OOOO   | OOOO\|O\|OOOO   | OOOO\|O\|OOOO   |
| Image Quality: | OOOO\|O\|OOOO   | OOOO\|O\|OOOO   | OOOO\|O\|OOOO   | OOOO\|O\|OOOO   |

9

67



\end{document}